\documentclass[aps,prfluids,nofootinbib,notitlepage,longbibliography,superscriptaddress,11pt,floatfix]{revtex4-2}
\usepackage{graphicx}
\usepackage[dvipsnames]{xcolor}
\usepackage{amsmath, amssymb,epstopdf}

\newcommand{\mylab}[3]{\raisebox{#2}[0mm][0mm]%
           {\makebox[0mm][l]{\hspace*{#1}{#3}}}}
\newcommand\Rey{\mbox{\textit{Re}}}

\newcommand{\dotted}{\protect\mbox{${\mathinner{\cdotp\cdotp\cdotp\cdotp\cdotp\cdotp}}$}}
\newcommand{\dashed}{\protect\mbox{-\ -\ -\ -}}

\newcommand{\full}{\protect\mbox{------}}
\def\spacce#1{\hskip #1pt}
\def\drawline#1#2{\raise 2.5pt\vbox{\hrule width #1pt height #2pt}}
\def\solid{\drawline{24}{.5}\nobreak}

\def\bdash{\hbox{\drawline{5.8}{.5}\spacce{2}}}

\def\dashed{\bdash\bdash\bdash\nobreak}

\def\bdot{\hbox{\drawline{1}{.5}\spacce{2}}}

\def\dotted{\hbox{\leaders\bdot\hskip 24pt}\nobreak}

\def\chndot{\hbox%
{\drawline{4.6}{.5}\spacce{2}\drawline{1}{.5}\spacce{2}\drawline{4.6}{.5}\spacce{2}\drawline{1}{.5}\spacce{2}\drawline{4.6}{.5}}\nobreak}

\def\dd{{\, \rm{d}}}

\def\bra{\langle}
\def\ket{\rangle}
\def\p{\partial}

\def\beq{\begin{equation}}
\def\eeq{\end{equation}}
\def\la{\label}
\def\ii{{\rm i}}

\def\degree{{$^{\rm o}$}}
\def\Real{{\rm Re}}
\def\r#1{(\ref{#1})}

\def\bu{\boldsymbol{u}}
\def\bv{\boldsymbol{v}}
\def\bw{\boldsymbol{w}}
\def\bE{\boldsymbol{E}}

\def\bm{\boldsymbol{m}}
\def\tL{\widetilde{L}}
\def\hp{\widehat{p}}
\def\hu{\widehat{u}}
\def\hw{\widehat{w}}

\colorlet{refcolor1}{black}
\colorlet{refcolor2}{black}

\begin{document}
\title{Logarithmic-layer turbulence: a view from the wall}

\author{Miguel P. Encinar}
\email[]{mencinar@torroja.dmt.upm.es}
\affiliation{School of Aeronautics, Universidad Polit\'ecnica de Madrid, 28040 Madrid, Spain}
\author{Javier Jim\'enez}
\affiliation{School of Aeronautics, Universidad Polit\'ecnica de Madrid, 28040 Madrid, Spain}

\date{\today}

\begin{abstract}
  The observability of the flow away from the wall in turbulent channels is studied using noiseless,
  although potentially incomplete, wall measurements. Reconstructions of the
  velocities are generated using linear stochastic estimation. All the velocities are well
  reconstructed in the buffer layer, but only relatively large `wall-attached' eddies are
  observable farther from the wall. In particular, large-scale structures that account for
  approximately 50\% of the total kinetic energy and of the tangential Reynolds stress are
  accurately captured up to $y/h \lesssim 0.2$. It is argued that this should allow the use of
  wall sensors for the detection (and potential control) of the large eddies populating the
  logarithmic region, and that it suggests a natural quantitative definition for
  `wall-attached' eddies.
\end{abstract}

\maketitle

\section{Introduction}\label{sec:intro}

This paper deals with the possibility of observing the flow far from the wall in a turbulent
channel, using only measurements at the wall. There are at least two reasons why this may be
desirable. The first one is purely scientific, because the relation between the near-wall
and outer flows has been one of the most controversial questions in wall-bounded turbulence,
and it stands to reason that influence and observability are related. If A does not
influence B, it is unlikely that the former can be reconstructed from observations of the
latter, and, conversely, if A can be reconstructed from B, it is unlikely that the two are
completely unrelated. Note that the inverse is not true, especially if, as in this paper, we
restrict ourselves to linear reconstructions. If A cannot be reconstructed from B, it cannot
be concluded that the two are independent. In addition, reconstruction by itself does not give
information on the direction of causality, if any. That B reconstructs A does not imply that
B causes A, or the other way around. We focus on variables that can be reconstructed from
wall measurements, and, in that regard, we will be able to propose criteria for which
variables are `attached' at what scales, in the sense that their influence extends to the wall, although
these criteria might not coincide with Townsend's dynamical definition of attached
structures \cite{tow:61}. On the other hand, although it is probable that variables that
cannot be reconstructed from the wall are `detached', in the sense of having little or no
influence over it, it follows from the previous discussion that this cannot be guaranteed.

A second reason for our interest in reconstruction is the possibility of active control of wall
turbulence. This is a classic goal of turbulence research, whether to decrease or increase
wall friction, reduce noise or other applications. Transportation of fluids through pipes,
such as oil, gas or water, would benefit from drag reduction, and so would vehicles moving
in fluids. In other engineering applications, such as combustion
or heat transfer, or even aerodynamics, it may be desirable to increase the turbulent
intensities close to walls to promote mixing or to delay separation. There is a well-developed
theory for the optimal control of linear systems, and, although turbulence is nonlinear,
there is extensive evidence that a substantial fraction of the dynamics of shear-driven
flows can be linearised \citep{jim13_lin}. This has led to heuristic and theoretical
active-control schemes that manipulate friction in turbulent channels at moderate Reynolds
numbers, at least in simulations \citep{choi94,far_ioa96con}, but they are not free from
problems and ambiguities.

\begin{figure}
\vspace*{2ex}%
\centering
\includegraphics[scale=1.23]{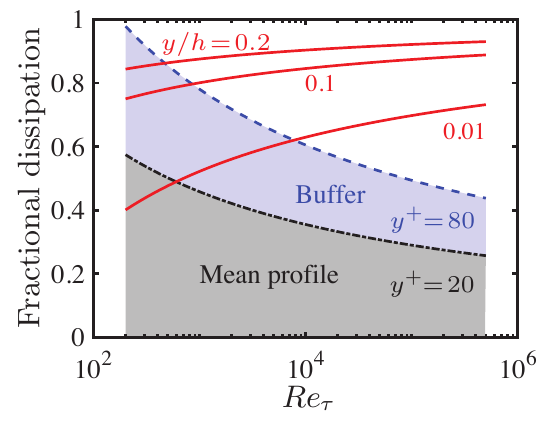}%
\mylab{-0.44\textwidth}{0.26\textwidth}{(a)}%
\hspace*{2mm}%
\raisebox{-3pt}[0pt][0pt]{\includegraphics[height=0.325\textwidth,clip]{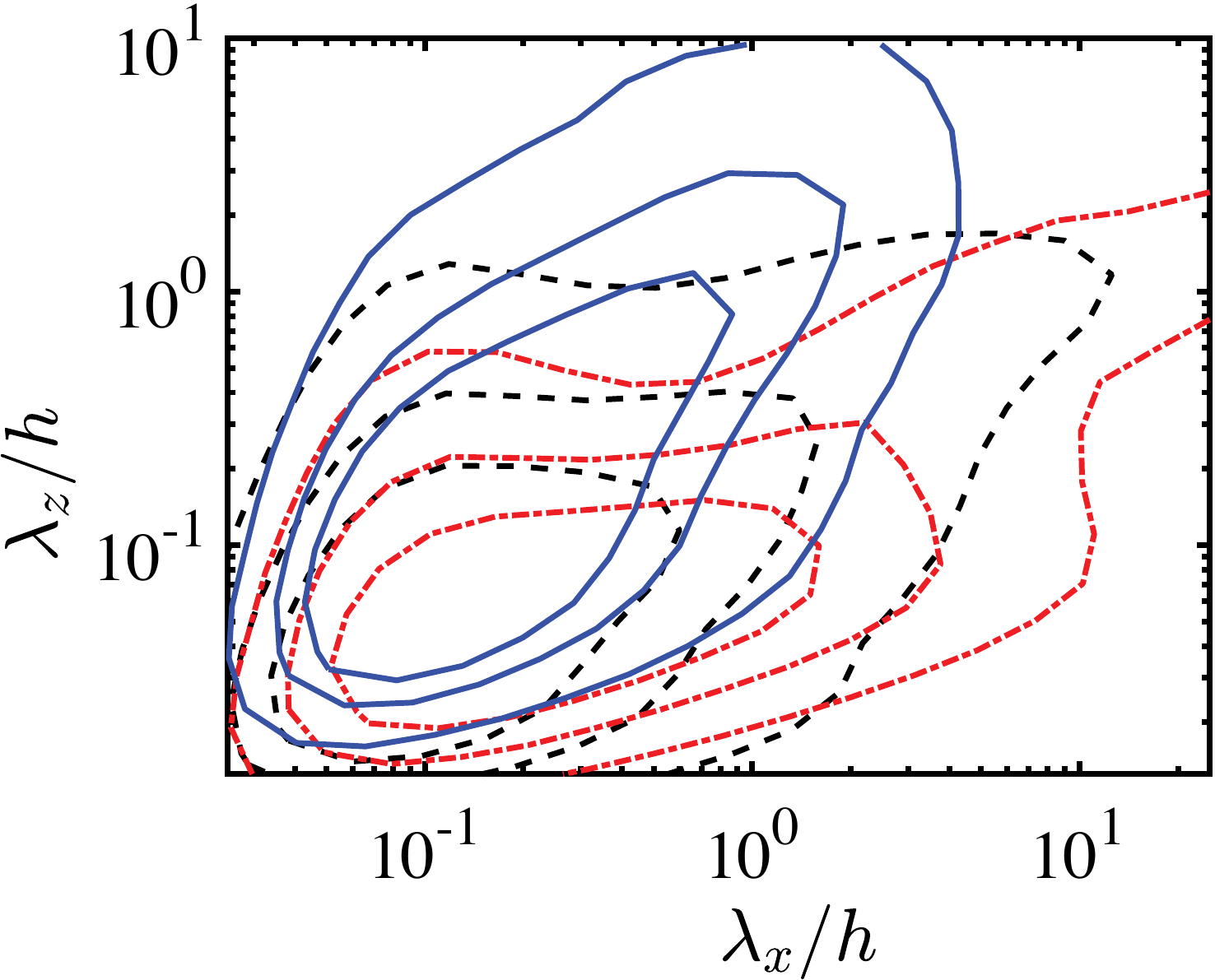}}%
\mylab{-0.39\textwidth}{0.26\textwidth}{(b)}%
\caption{%
(a) Fraction of the total energy dissipation below a given distance from the wall, as a
function of the Reynolds number. The dashed lines are scaled in wall units, and the labels
refer to the predominant contributor to the dissipation in each region. The solid lines are
three distances from the wall, scaled in outer units \citep[adapted from][]{jim18}. 
\textcolor{refcolor2}{(b) Premultiplied spectra of \cite{hoy06} at $y^+=5$; \chndot, streamwise stress; \dashed, spanwise stress; \solid, pressure. The contours of each spectrum contain 50\%, 70\% and 90\% of its spectral
mass. $\lambda_x$ and $\lambda_z$ are respectively the streamwise and spanwise wavelengths.}
}
\label{fig:disip}
\end{figure}

The first problem has to do with the scales involved. Most of the dissipation in
wall-bounded flows is contained within or below the logarithmic layer (see figure
\ref{fig:disip}a), suggesting that effective control should act near the wall. As a
consequence, most schemes have targeted the buffer-layer, which contains the strongest
turbulent fluctuations, but technological considerations suggest that practical applications
should rather centre on the logarithmic-layer eddies. Thus, for a flow of thickness $h$ and
Reynolds number $Re_\tau=u_\tau h/\nu$, where the `+' superscript denotes wall units based on
the friction velocity $u_\tau$ and on the kinematic viscosity $\nu$, the sizes and passing
times of the structures in the buffer layer decrease proportionally to $Re_\tau$ when expressed
in outer units. On the contrary, when the distance, $y$, from the wall is a fixed fraction
of $h$, sizes and times scale in outer units, or at most depend logarithmically on the
Reynolds number. For example, in a water pipe with $h\approx 1$\,m and bulk velocity
$U\approx 1$\,m/s $(Re_\tau\approx 10^5)$, the wall-normal velocity eddies in the buffer-layer
have length and time scales of $\Delta x\approx 1$\,mm $(\Delta x^+=100)$ and $\Delta
t\approx 0.1$\,ms, but those at a distance from the wall $y/h\approx 0.1$ are $\Delta
x\approx 20$\,cm and $\Delta t \approx 0.2$\,s, where $\Delta x$ is the streamwise length.
 Similar differences apply to the flow over
aeroplane wings.


However, moving away from the wall is not without cost. Figure \ref{fig:disip}(a) presents
the fraction of the total dissipation below a given distance from the wall as a
function of the Reynolds number \cite{jim18}. An `ideal' control strategy would probably
result on the complete elimination of the turbulent energy dissipation over some part of the channel.
Below $y^+ = 20$, most of the dissipation is due to the mean shear,
and is independent of fluctuations. Centring on the remaining dissipation, in the example above, the velocity fluctuations in the
logarithmic region, defined as $80\nu/u_\tau \le y \le 0.2h$, are responsible for about 40\%
of the energy loss, but the scales in its lower limit are those discussed above as being too
small. If we want to restrict ourselves to fixed fractions of the flow thickness, to avoid
wall scaling, the maximum fraction of the drag that could potentially be saved by completely
removing all the dissipation between $y/h=0.01$ and 0.2 would be approximately 25\% (at $Re_\tau \approx 10^5$).
Decreasing the lower limit of wall distances increases the potential gain, but at the cost
of having to deal with smaller length and time scales. Raising the upper limit is less
effective. These estimations are only upper bounds, because turbulent dissipation can
probably not be completely damped, but they show that the logarithmic layer is a `preferred
region' in which some control authority still remains, while the size of the eddies stays
bounded from below as the Reynolds number increases.

A second problem is that most active-control schemes assume knowledge of the flow at wall
distances of the order of the size of the structures to be controlled. Typically, the
assumption is that the flow is fully known, as in direct numerical simulation (DNS), or at
least that a two-dimensional section of some variables is available, as in particle-image
velocimetry. Unfortunately, most practical observations are limited to the wall, where the
only accessible variables are the pressure and the two shears. This problem is especially
acute when we are interested in the logarithmic layer, which is separated from the wall by
the very active buffer layer. Figure \ref{fig:disip}(b) shows spectra of the three
observable variables at the wall. All of them are dominated by a core whose position
scales in wall units, but they also include a larger-scale component, which is long for the two
shears and wide for the pressure, with wavelengths that scale proportionally to $h$, and which are 
associated with eddies farther from the wall \cite{hoy06}.

This paper focuses on the properties of the estimations of logarithmic-layer turbulence using
noiseless, although potentially incomplete, wall measurements from DNS. The widespread
availability of data in the last years has allowed various research groups to test the
performance of similar methods with considerable success
\cite{baa:hut:16,baa:hut:17,ben:yeg:17,ill:mon:18,sas:vin:19}. These works differ from ours in that
most of them focus on the use of physically motivated linear models \cite{ben:yeg:17,ill:mon:18} to relate velocity
measurements at one $y$ to predict the same velocity component at another wall-normal
distance. Others \cite{milano2002neural,sas:vin:19}, also explore the performance of nonlinear models, with modest gains over their proposed linear method. In contrast, our methodology is linear regression, which has been successfully
used for conditional flow reconstruction away from walls \citep{adr:moi:88}, or to
synthesise buffer-layer velocities from experimental measurements farther away
\cite{baa:hut:16}. This `linear stochastic estimation' (LSE), although apparently naive, is
statistically optimal for linear systems, and we will show that it is adequate to
reconstruct the Reynolds-stress structures in the logarithmic layer, using only variables
that can be measured at the wall. A closely related investigation is \cite{VilaFlores:2018}, which 
analyses the correlation between wall pressure at the wall-normal velocity farther into the flow, but, 
to our knowledge the present paper is the first to one to use the full set of wall observables to 
reconstruct the full internal flow field, including the relative importance of the different observables 
on the three velocity components, and the identification of the accessible scales.   
Finally, it should be noted that, although not specifically addressed in the present work,
similar methods should apply to transitional flows, where scales are typically larger, the
role of linearity is more pronounced, and are thus more amenable to active control.

The remainder of this paper is organised as follows. The numerical
datasets and the details of the reconstruction algorithm are presented in
\S\ref{sec:experiments} and \S\ref{sec:methods}, respectively. Section \ref{sec:results}
discusses the properties of the reconstruction, and \S\ref{sec:conclusions} concludes. 
Some preliminary work on the reconstruction of vorticity using similar techniques
was performed at the 2018 CTR Stanford Summer Program \cite{encinar:ctr18}, and an
early partial version of the present manuscript is \cite{recon:arxiv}.

\section{Numerical experiments}\label{sec:experiments}


We consider incompressible flow in channels whose half-height is $h$. The streamwise,
wall-normal and spanwise directions are $x, y, z,$ respectively, and $u, v, w, p$ are the
fluctuations of the corresponding velocity components and of the kinematic pressure.
Capitals are used for averaged quantities, and primes for standard deviations. The
computational domain is periodic in both wall-parallel directions, with periods $L_x$ and
$L_z$. The spatial discretisation is Fourier spectral in those directions, dealiased using
the $2/3$ rule, and compact finite differences or Chebyshev expansions in the
wall-normal direction, depending on the Reynolds number. Table \ref{tab:sim} collects basic
information on the simulations, and refers to the original publications for further
details.

\begin{table}
\caption{Parameters of the simulations. $L_{x,z}$ are the streamwise and spanwise period of
the numerical box, and $N_{x,y,z}$ are the number of collocation points of the stored grid
in each direction.  \textcolor{refcolor2}{$\Delta_{x}^+,\Delta_{z}^+$ are the stream and spanwise grid resolutions of the DNS. $\Delta_{y_{\max(\min)}}^+$ is te maximum (minimum) resolution across the wall-normal direction. $Tu_\tau/h$ is the time spanned by the aproximatelly equispaced snapshots in eddy turnovers.} $D_y$ represents the discretisation used for the wall-normal direction:
`CH' refers to Chebyshev polynomials and `FD' to compact finite differences. $N_f$ is the
number of fields used to compute statistics.}
    \label{tab:sim}
  \begin{center}
    \def~{\hphantom{0}}
    \begin{ruledtabular}
    \begin{tabular}{lcccccccccccc} 
      Name  & $\Rey_\tau$ & $(L_x, L_z)/h$ & $N_x$ & $N_y$ & $N_z$ & $\Delta x^+$ & $\Delta z^+$ & $\Delta y^+_{\max{}(\min{\!})}$ & $Tu_\tau/h$ & $D_y$ & $N_f$ & Reference\\\hline
      S1000 & 932 & $(2\pi, \pi)$ & $512$ & $385$ & $512$ & 12 & 6& 7.7(0.03) & 18 & CH & $1800$ & {\small \cite{lozano-time}}\\ 
      S2000 & 2003 & $(2\pi,\pi)$ & $1024$ & $633$ & $1024$ & 12 & 6& 8.9(0.3)& 11 & FD & $800$ & {\small \cite{lozano-time}}\\
      F2000 & 2003 & $(8\pi, 3\pi)$ & $512$ & $512$ & $512$ & 24 & 9& 10.6(1.7) & 14 &FD & $1200$ & {\small \cite{encinar19}}\\
      F5300 & 5300 & $(8\pi, 3\pi)$ & $1024$ & $1024$ & $768$ & 22 & 12& 14(1.27) & 30 &FD & $3000$ & {\small \cite{encinar19}}\\
    \end{tabular}
    \end{ruledtabular}
  \end{center}
\end{table}

Since our focus is on the logarithmic layer, and on scales that are large compared to the
viscous length, the spatial resolution of the database is not critical except very near the
wall, but the longest and widest scales, particularly those of the streamwise velocity,
require large boxes and long integration times to be accurately represented. To complement
the high-resolution, but relatively small, boxes of S1000 and S2000 while keeping the
computations economical in terms of storage, we use the large-box simulations F2000 and
F5300. They are computed as DNSes but stored at the resolution of large-eddy simulations,
$\delta x^+ \approx 120$ and $\delta z^+ \approx 90$, which is still sufficient to study the
energy- and stress-carrying eddies in the logarithmic layer. The pressure at the wall in these data sets is computed
retaining the effect of the discarded scales \cite{encinar19}.

\section{Linear Stochastic Estimation}\label{sec:methods}

Stochastic estimation is the approximation of an unknown, $\boldsymbol{u}$, using the
statistical information from an observable, $\boldsymbol{E}$. To fix ideas, we consider
$\boldsymbol{u}$ to be a vector with the streamwise velocity component at every grid point,
\begin{equation}
  \boldsymbol{u}(y) = [u_i(y)],
\label{eq:vectorize}
\end{equation}
where the $i$ subindex represents the two grid indices $i_x=1\ldots N_x$ and $i_z=1\ldots
N_z$. The wall-normal $(\bv)$ and spanwise $(\bw)$ velocity components are analogously
expressed, and any procedure described for $\boldsymbol{u}$ is equivalently applied to
them. \textcolor{refcolor2}{Owing to the linearity of the method, optimising the individual velocity components is equivalent to optimising the velocity vector as a whole.} Our observables are,
\beq
  \boldsymbol{E} = \left[p_i(0), \partial_y u_i(0), \partial_y w_i(0)\right],
\eeq
where indices are defined as above, but $\bE=[E_{i(s)}]$ includes a second index, $s=1\ldots
3$, representing the three observables. The best statistical estimation, $\bu^\dagger(y)$, of $\bu(y)$
under the $L_2$-norm, is the conditional average of $\bu(y)$ given $\bE$, i.e. $\langle
\boldsymbol{u}(y)|\boldsymbol{E}\rangle$, where $\bra\cdot\ket$ stands for ensemble
averaging. When expanded as a Taylor series in $\boldsymbol{E}$, and truncated at the linear
term, it is called a linear stochastic estimator (LSE), and is numerically equivalent to the
linear mean-square approximation of $\bu$ in terms of $\bE$ \citep{adrian94}. The estimator
$\tL_{ij(s)}(y)$ in
\begin{equation} 
u^\dagger_i (y) = \tL_{ij(s)}(y) E_{j(s)},
\label{eq:recon}
\end{equation} 
where repeated indices imply summation, is obtained by minimising
\begin{equation}
  \left\bra\left(u_i(y) - u_i^\dagger(y)\right)^2\right\ket=
  \left\bra\left(u_i(y) - \tL_{ij(s)}(y) E_{j(s)}\right)^2\right\ket,
\end{equation}
which reduces to solving the linear system,
\begin{equation}
\langle E_{j(s)}E_{m(r)} \rangle \tL_{ij(s)}(y) = \langle u_i(y) E_{m(r)}\rangle,
\label{eq:LSE}
\end{equation}
where $\langle E_{j(s)}E_{m(r)} \rangle$ is the autocorrelation tensor of the observables,
and $\langle u_i(y)E_{m(r)}\rangle$ is the cross-correlation between the unknowns and the
observations. The presence of the latter in the right-hand side of \r{eq:LSE} codifies the
intuition expressed in the introduction that two quantities are only mutually observable if
they are correlated. It also implies that reconstruction is a largely symmetric property. If
A allows us to estimate B, B should allow us to estimate A. Note that the only information
required to compute $\tL_{ij(s)}$ is contained in the two correlation tensors, and that
$\bu$ is only assumed to be linear with respect to $\bE$. There is no assumption about the
linearity of the velocity field in terms of the geometric coordinates.

\textcolor{refcolor1}{
Equation \eqref{eq:LSE} can be manipulated to increase or decrease the number of observables used to estimate the velocity field. A reduced estimator may be desirable to separate the contributions of the different observables to the full model, or for practical considerations. For example, if only the pressure ($s = 1$) is used to estimate $u$, \eqref{eq:LSE} becomes,
\begin{equation}
    \langle E_{j(1)}E_{m(1)} \rangle \tL_{ij,p}(y) = \langle u_i(y) E_{m(1)}\rangle.\label{eq:lse0}
\end{equation}
In the reduced system, $\langle E_{j(1)}E_{m(1)} \rangle$ is the autocorrelation of the pressure at the wall, and $\langle u_i(y) E_{m(1)}\rangle$ the cross-correlation of $p$ at the wall with $u$ at every wall distance. Note that $\tL_{ij,p}(y) \neq \tL_{ij(1)}(y)$, as the latter contains information about the shears through the inverse of $\langle E_{j(s)}E_{m(r)} \rangle$.
}

In our case, $\bE$ is a vector of length $3N_xN_z$, resulting in a large correlation matrix
that has to be inverted in order to solve for $\widetilde{L}_{ij}$ in \r{eq:LSE}. This can
be alleviated by exploiting the periodicity of the domain and projecting on the Fourier
basis. For every pair of Fourier modes, $\widehat{\bE} (k_z, k_x)$ and $\widehat{u}(k_z, y,
k_x)$, equation \eqref{eq:LSE} becomes,
\begin{equation}
\bra \widehat{E}_s\widehat{E}^*_r \ket (k_z, k_x) \widehat{L}_s(k_z,y, k_x) = 
\bra \widehat{u}(k_x,y,k_z)\widehat{E}^*_r(k_x, k_z)\ket,
\label{eq:SLSE}
\end{equation}
where the carat denotes Fourier transformation, $k_i$ are the wavenumbers, and the
asterisk is complex conjugation. The now more feasible calculation consists on solving
$N_zN_x$ linear problems of dimension three. This procedure is also known as spectral linear
stochastic estimation \citep[SLSE,][]{slse}. It is formally equivalent to LSE, but it
behaves better with incomplete or noisy data because it avoids the spurious correlations
between the orthogonal Fourier basis functions.
The inverse Fourier transform of $\widehat{L}_s$, as obtained from \eqref{eq:SLSE}, can be used
to reconstruct the velocity fields in physical space, and \eqref{eq:recon} becomes a
discrete convolution,
\begin{equation}
  {u^\dagger}(i_x, y,i_z) = 
  \sum_{\forall j_{x}, j_{z}} {\tL}_s(i_x - j_{x},y, i_z - j_{z}){E_s}(j_{x}, j_{z}),\label{eq:last_lse}
\end{equation}
where the full operator $\tL_{ij(s)}$ has been expressed as a block-Toeplitz matrix whose
blocks are shifted versions of the inverse transform $L_s(i_x,i_z)$ of the individual
Fourier estimators. In a mean-square sense, the operator $\tL_{ij(s)}(y)$ is by construction
the best possible data-driven linear operator for the reconstruction of the velocity
components at a given wall distance. This can be seen from \eqref{eq:last_lse}, or from its
continuous equivalent
\beq
  {u^\dagger}(x,y,z) =\iint \tL_s (x-x', y, z-z') E_s(x',z') \dd x'\dd z' ,
  \la{eq:cont_lse}
\eeq
because it uses all the contemporaneous measurements at the wall to reconstruct the velocity
at each point.

The accuracy of the reconstruction can be improved by enriching $\boldsymbol{E}$
with information from previous time steps. However, some experimentation with physically
motivated time sampling resulted in a best-case reduction of 14\% in the reconstruction error
for S1000, compared with only using instantaneous data, at a cost between 8 and 16 times higher.
As a consequence, these experiments were not pursued, and the results in this paper only use
contemporaneous observables.

\section{Results}\label{sec:results}

\begin{figure}
\vspace*{3ex}%
  \centering
\includegraphics[width=\textwidth]{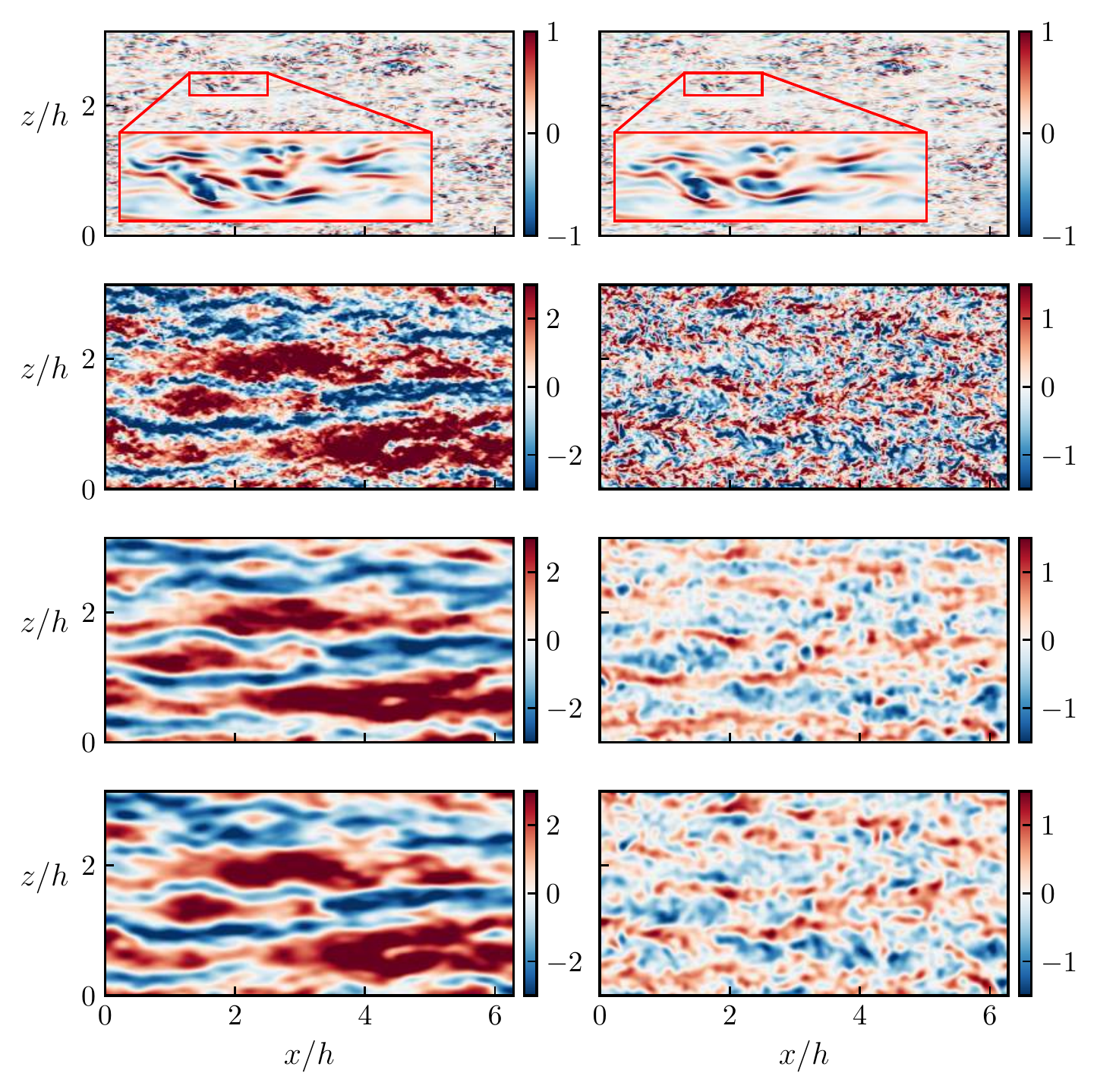}
\mylab{-0.74\textwidth}{0.98\textwidth}{(a)}%
\mylab{-0.30\textwidth}{0.98\textwidth}{(b)}%
\mylab{-0.74\textwidth}{0.75\textwidth}{(c)}%
\mylab{-0.30\textwidth}{0.75\textwidth}{(d)}%
\mylab{-0.74\textwidth}{0.52\textwidth}{(e)}%
\mylab{-0.30\textwidth}{0.52\textwidth}{(f)}%
\mylab{-0.74\textwidth}{0.29\textwidth}{(g)}%
\mylab{-0.30\textwidth}{0.29\textwidth}{(h)}%
\caption{(a,b) Snapshots of S1000 at $y^+= 10$.
(a) $v^+$.
(b) $(v^\dagger)^+$.
(c-h) F5300 at $y/h = 0.1$.
(c) $u^+$. (d) $v^+$.
(e) $(u^\dagger)^+$. (f) $(v^\dagger)^+$.
\textcolor{refcolor1}{(g)} $u_G^+$, filtered with \r{eq:G} and $\Delta_x \times\Delta_z = (4\times 2)y$.
\textcolor{refcolor1}{(h)} $v_G^+$ with $(2\times 2)y$.
}
\label{fig:planes}
\end{figure}

We use the methodology just described to generate flow reconstructions for all the cases in
table \ref{tab:sim}. \textcolor{refcolor2}{First, we tested if reconstructing fields `in-sample' made a difference over reconstructing fields `out-of-sample'. For this purpose, we computed the operators for S1000 and F2000 using only half of the available velocitity fields. The correlations can be symmetrised, effectively multipliying by four the number of independent samples \cite{moi:mos:89}. Thus, the linear operator converges reasonably fast and, no difference can be found between the in-sample and out-of-sample snapshots. Considering the outcome of this experiment, we use all the flow fields available to compute the correlations in \eqref{eq:SLSE}.}

Figure \ref{fig:planes}(a,b) shows snapshots of the true and
reconstructed wall-normal velocity very close to the wall $(y^+ \approx 10)$ in S1000. The
reconstructed field captures the original in almost full detail, and the same is true for
the other two velocity components, and for the Reynolds product $u^\dagger v^\dagger$ (not shown). 

A similar level of accuracy only holds for $y^+ \lesssim 20$. Figures \ref{fig:planes}(c-h)
compare the true $u$ and $v$ with their reconstructions in the logarithmic layer $(y/h =
0.1,\,y^+ \approx 500)$. At this distance from the wall, most eddies with sizes smaller than $O(y)$ are
missing from the reconstructions in figure \ref{fig:planes}(e,f). This agrees with the idea
that `detached' structures do not leave a footprint at the wall, and cannot be reconstructed
from wall information. However, it is clear from the comparison of figures
\ref{fig:planes}(e,f) and \ref{fig:planes}(c,d) that some large-scale information survives
the reconstruction. This is best seen by comparing the reconstructed fields with those in
figure \ref{fig:planes}(g,h), where the true velocities have been filtered with a low-pass
Gaussian kernel,
\begin{equation}
  G(x, y, z) = \left(\frac{2\pi}{\Delta_x\Delta_z}\right)^{1/2}\,
  \exp\left[ -(\pi x/\Delta_x)^2 -(\pi z/\Delta_z)^2 \right],
  \label{eq:G}
\end{equation}
where the filter widths are empirically adjusted to retain those large scales for which the
error in the reconstruction is less than $50\%$. Figure \ref{fig:planes} shows that, even
far from the wall, the wall-normal velocity is organised into streaks. The reconstructed
velocity fields include streaks of $u$, which account for most of the kinetic energy,
and weaker $v$ structures of similar size. The energy of the latter is only about 10\% of
the total $v^2$-energy at that height, but their correlation with $u$ is strong enough for
the reconstructed velocity field at the height of figure \ref{fig:planes}(c-h) to contain
approximately $52\%$ of the tangential Reynolds stress. \textcolor{refcolor2}{The filtered fields hold a similar amount of Reynolds stress, $45\%$, reinforcing the visual confirmation that the filter widths used are reasonable. The fair amount of Reynolds stress captured is not surprising if we consider that the spectral structure
parameter,} $-\phi_{uv}/(\phi_{uu}\phi_{vv})^{1/2}$, is known to be very close to unity for those
large modes \citep{jim:ala:flo:04}. The long $v$ structures seen in the figure are those
responsible for `lifting' and maintaining the streaks \citep{kli:rey:sch:run:67}.

\begin{figure}
  \centering
  \includegraphics[width=.88\textwidth]{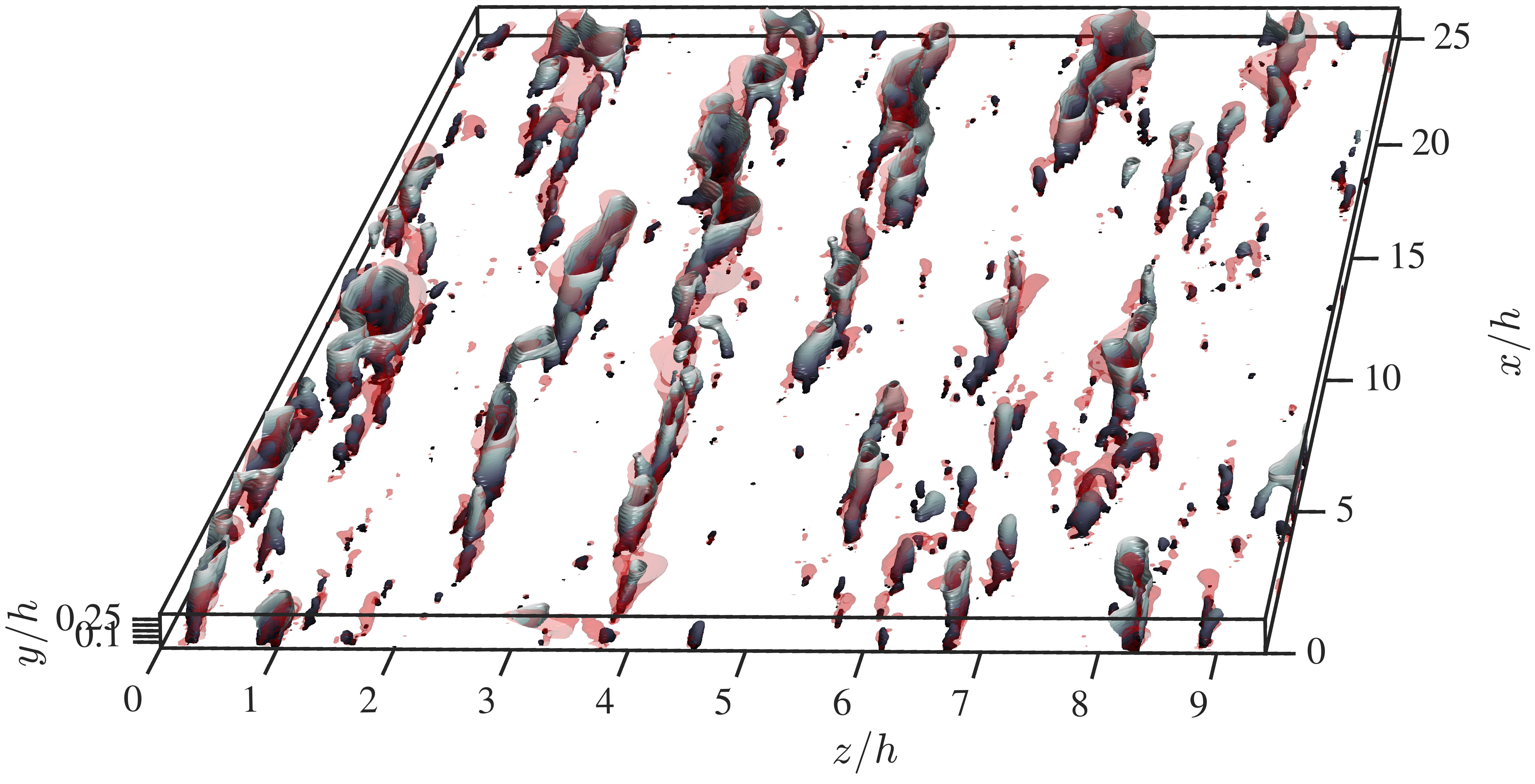}%
  \mylab{-0.74\textwidth}{0.4\textwidth}{(a)}
  \includegraphics[width=.88\textwidth]{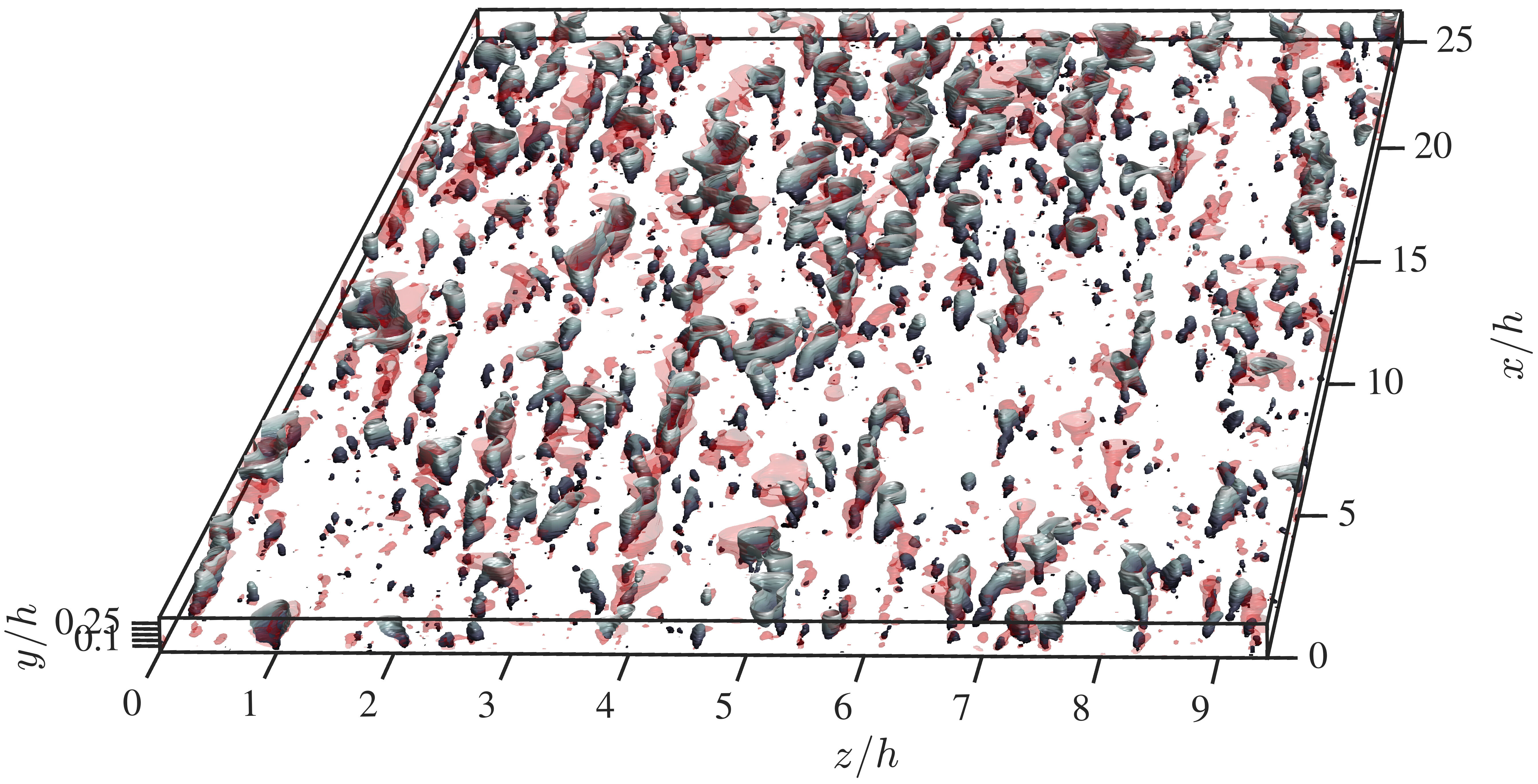}%
  \mylab{-0.74\textwidth}{0.4\textwidth}{(b)}

  \caption{\textcolor{refcolor2}{Snapshot of the F5300 reconstruction. (a) The solid grey isosurfaces are $u^\dagger/u^{\dagger\prime}(y) < -1.5$, coloured with the distance from the wall, and the translucent red ones are $u_G/u_G^\prime(y) < -1.5$. (b) The solid grey isosurfaces are $u^\dagger v^\dagger/(u^\dagger v^\dagger)^\prime(y) < -1.5$, coloured with the distance from the wall, and the translucent red ones are $u_Gv_G/(u_Gv_G)^\prime(y) < -1.5$.
  The real velocities $u$ and $v$ (but not $u^\dagger$ and $v^\dagger$) are filtered using \eqref{eq:G} with $\Delta_x \times\Delta_z = (4\times 2)y$ and $\Delta_x \times\Delta_z = (2\times 2)y$, respectively.}
  }\label{fig:streaks}
\end{figure}

\textcolor{refcolor2}{
The relation between the two velocity components is even clearer in figure
\ref{fig:streaks}, which shows an instantaneous three-dimensional representation of the
reconstructed velocity field below $y/h=0.25$. The solid grey isosurfaces of figure \ref{fig:streaks}(a) represent the
low-speed streaks of $u^\dagger$, while the red translucent ones are those of $u$. The original field has been filtered with \r{eq:G} to retain only
reconstructible scales. It is clear that, except for some discrepancies in the smaller
structures far from the wall, all of the large features of $u$, including short or `broken'
streaks, are well reproduced. Most interesting are the isosurfaces of $-u^\dagger v^\dagger$,
represented in solid grey in figure \ref{fig:streaks}(b). As mentioned above, the reconstruction of $v$ is poor at this
wall distance, but it is clear from the figure that the $v$-structures that can be
reconstructed are those associated with intense Reynolds stress. The isosurfaces of in the figure are the `ejections' and `sweeps' \citep[$Q_{2-4}$,][]{wal:eck:bro:72, wil:lu:72}, which are known
to mostly occur within the streamwise velocity streaks, and this arrangement is well captured by the
reconstruction. Except for several `detached' structures from the wall, most of the filtered Reynolds stress $Q_{2-4}$ structures that extend from the wall are well captured, and they are known to carry half of the total Reynolds stress \citep{loz:jim:2014}.
}

\begin{figure}[t]
  \centering
  \includegraphics[scale=1]{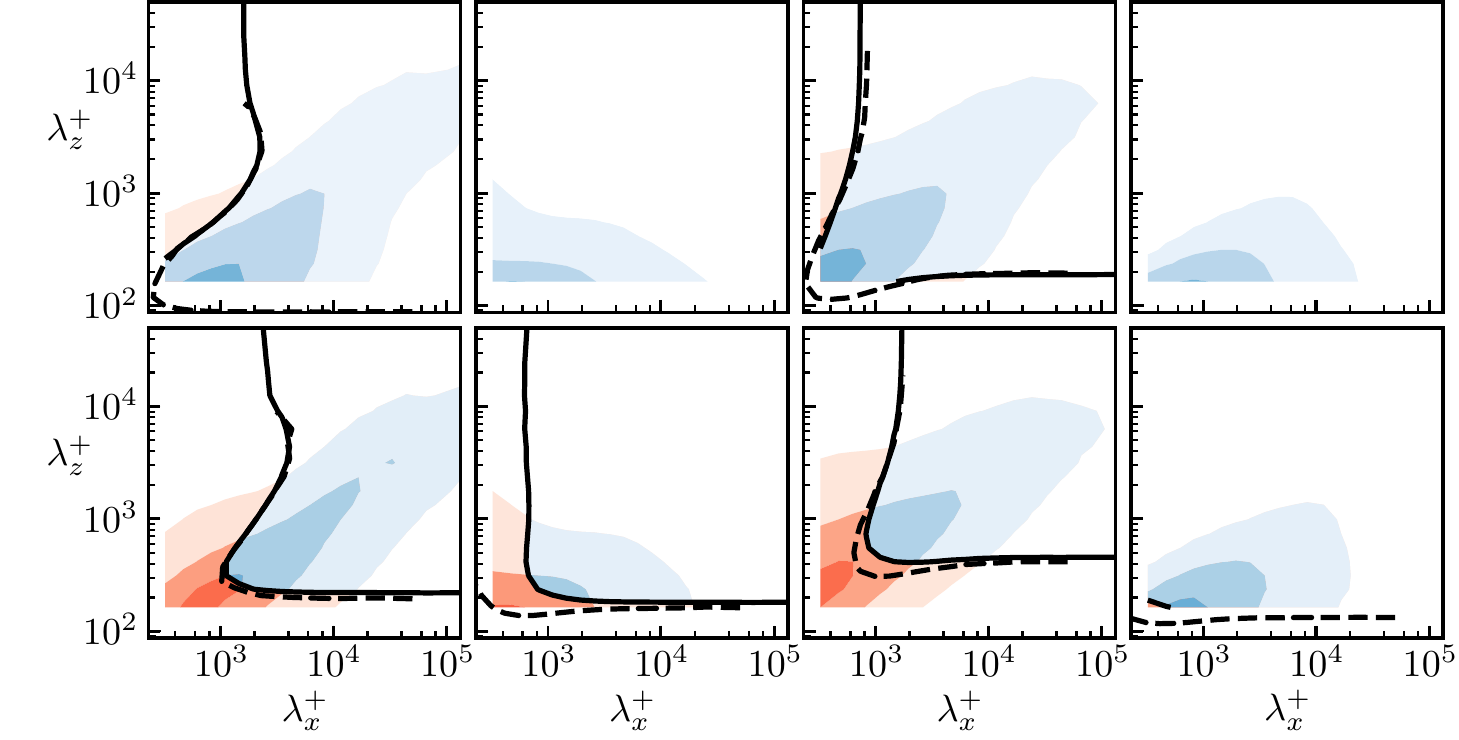}\\[10pt]%
    \includegraphics[scale=1]{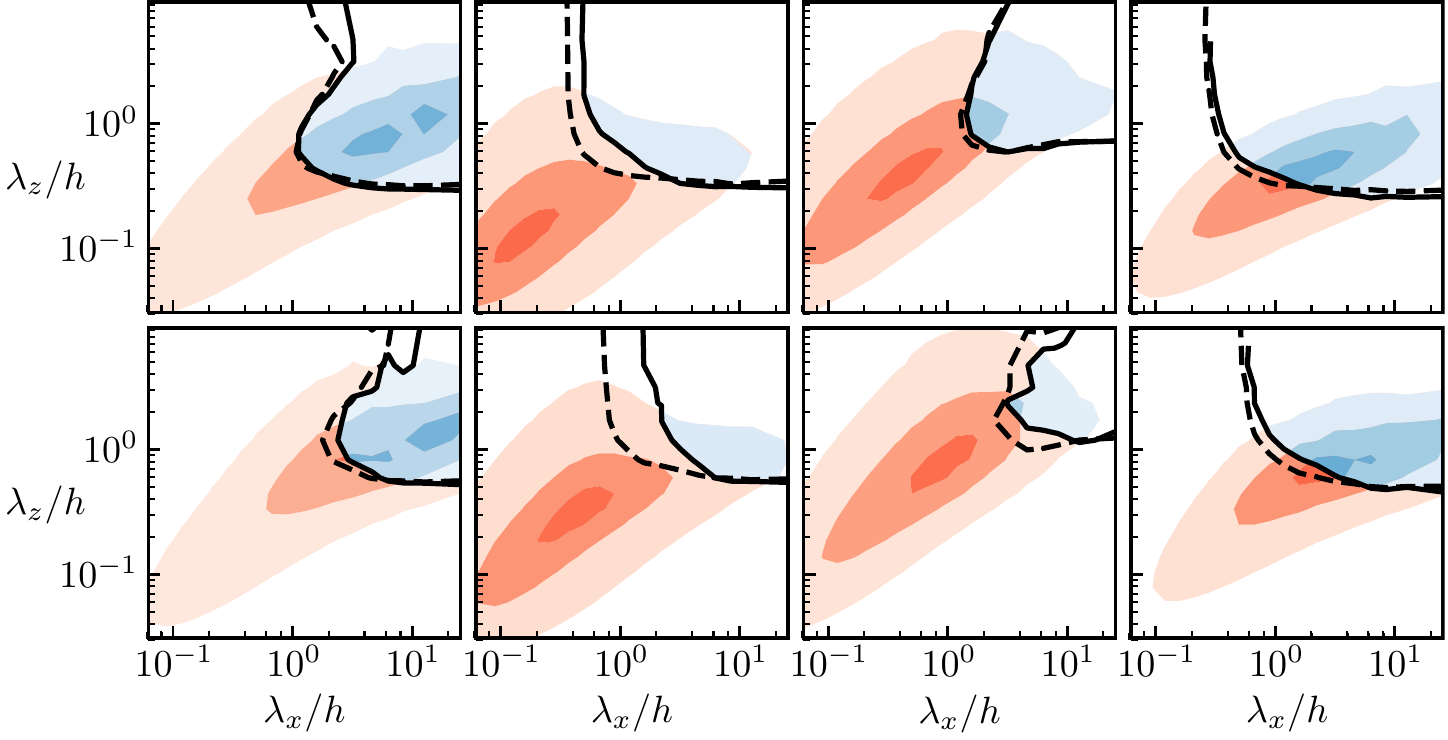}%
\caption{Spectral reconstruction error at different heights and Reynolds numbers, on top of
the premultiplied velocity spectra  and cospectrum(shaded) for F5300. The line contours are $R_{ab} = 0.5$:
\dashed F2000; \solid, F5300. The shaded contours contain $90\%, 50\%$ and $10\%$ of the
total energy or of the tangential Reynolds stress. From left to right, $k_xk_z\phi_{uu}$, $k_xk_z\phi_{vv}$, $k_xk_z\phi_{ww}$
and $-k_xk_z\phi_{uv}$. From top to bottom $y^+\approx 20$, $y^+ \approx 40$, $y/h = 0.1$, $y/h =
0.2$. The blue-shaded portion of the spectra marks the region reproduced with less than
$50\%$ error.
}
  \label{fig:spectra}
  \end{figure}

\begin{figure}[t]
  \centering
  %
  \includegraphics[scale=0.8]{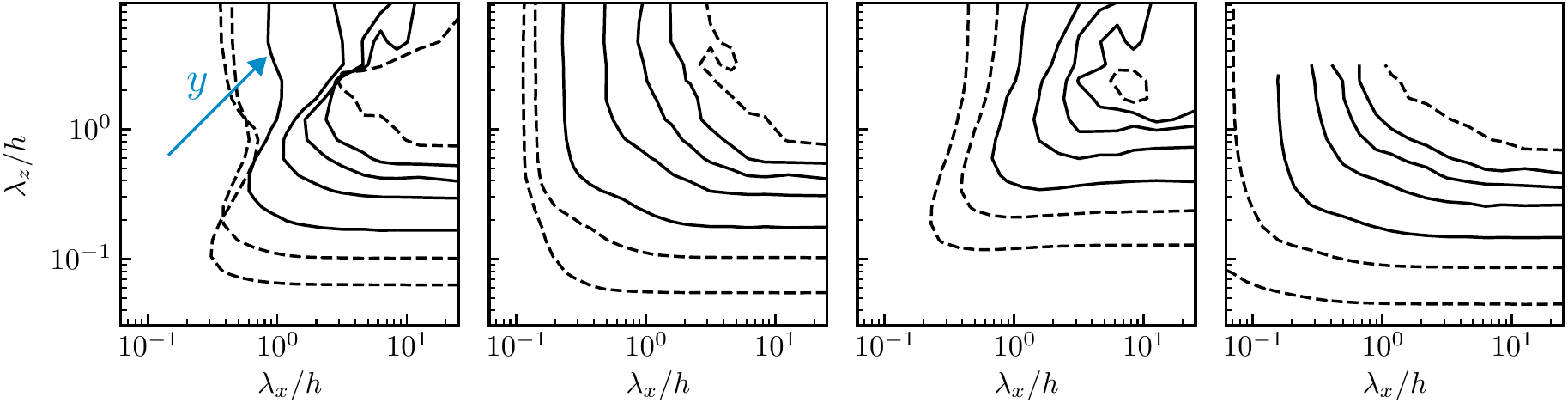}%
  \mylab{-12.4cm}{4cm}{(a)}%
  \mylab{-8.9cm}{4cm}{(b)}%
  \mylab{-5.4cm}{4cm}{(c)}%
  \mylab{-1.9cm}{4cm}{(d)}%
  \vspace{15pt}
  \includegraphics[scale=0.8]{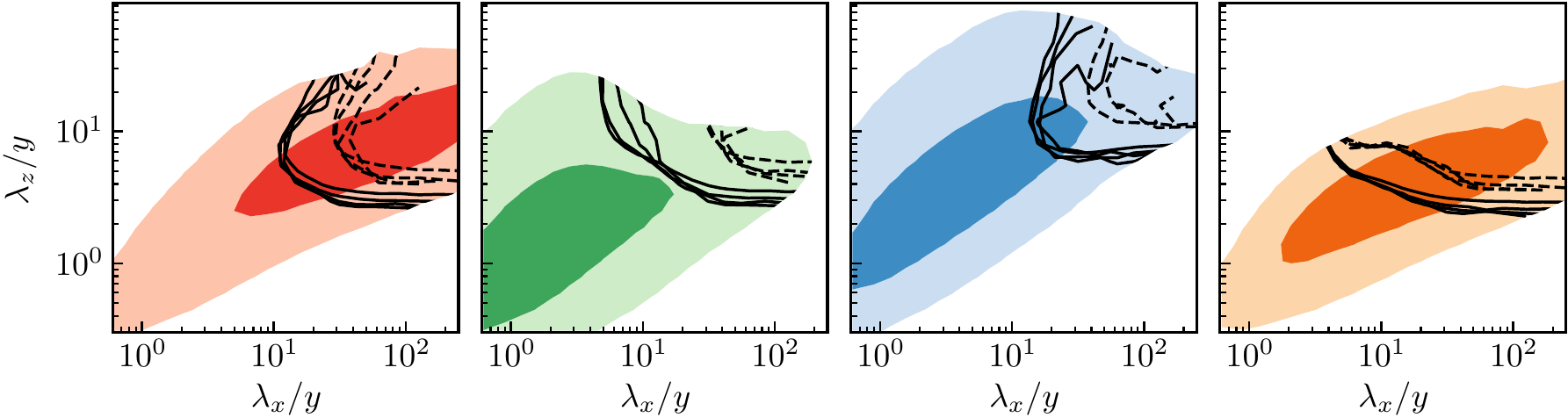}%
  \mylab{-12.4cm}{4cm}{(e)}%
  \mylab{-8.9cm}{4cm}{(f)}%
  \mylab{-5.4cm}{4cm}{(g)}%
  \mylab{-1.9cm}{4cm}{(h)}%
\caption{Scaling of the error isolines with the distance from the wall for F5300. From left
to right, $R_{uu}$, $R_{vv}$, $R_{ww}$, $R_{uv}$.
(a-d) Wavelengths scaled with the channel height. Contours are $R_{ab} = 0.5$ at $y/h =
[0.012, 0.025, 0.05, 0.1, 0.15, 0.2, 0.3]$. The solid lines are within the logarithmic
region, $150 \nu/u_\tau < y \leq 0.2 h$.
(e-h) Wavelengths scaled with the distance from the wall, including only data within the
logarithmic layer.  Solid isolines, $R_{ab} = 0.5$;  dashed, $R_{ab} = 0.25$; 
Shaded contours are the premultiplied spectra and cospectrum at $y/h = 0.1$.
}
\label{fig:spectra_sca}
\end{figure}

To quantify the performance of the reconstruction as a function of eddy size, we define the
fractional spectral error
\begin{equation}
R_{ab}(k_x, y, k_z) = \dfrac{\Real\left\bra%
\left(a-a^\dagger\right)\left(b-b^\dagger\right)^{*}\right\ket(k_x, y, k_z)}%
{\Real\bra a b^*\ket(k_x, y, k_z)},
\la{eq:R}
\end{equation}
where $a$ and $b$ stand for either $u$, $v$ or $w$. \textcolor{refcolor1}{The spectral error as defined above is related to other measures of correlation or `coherence' across wall distances. Other works using LSE evaluate its performance using the Linear Coherence Spectrum \cite{slse,baa:hut:16,mad:ill:19},
\begin{equation}
    \gamma^2_{ab}(k_x, y, k_z) = \dfrac{\Real\left\bra a^\dagger (b^\dagger)^*\right\ket(k_x, y, k_z)}{\Real\bra a b^*\ket(k_x, y, k_z)},
\end{equation}
which is related to $R_{ab}$ by,
\begin{equation}
    R_{ab} = 1 + \frac{\Real\langle a^\dagger(b^\dagger)^*\rangle}{\Real\langle ab^* \rangle} - 2\gamma^2.\label{eq:errga}
\end{equation}
It follows from \eqref{eq:errga} that the greater the coherence, the lower the reconstruction error. For the limiting cases where the reconstruction error $R_{ab}$ is either unity or zero, both quantities are complementary, i.e. if the coherence is one, the error is zero and \emph{vice versa}. One potential advantage of the spectral error is that the second factor in \eqref{eq:errga} penalises reconstructed energy without coherence, which may be significant for methods that are not $L_2$-optimal.
}

The contour of $R_{ab} = 0.5$ is plotted
at four wall distances in figure \ref{fig:spectra}, compared with the energy spectrum of the
three velocity components, and with the cospectrum of the tangential Reynolds stress. The
regions shaded in blue and red indicate $R_{ab} < 0.5$ and $R_{ab} > 0.5$, respectively. In
the first row, corresponding to $y^+ = 20$, the low-error region encloses most of the
spectral plane and accounts for most of the energy, especially for $v$. As we saw in figure
\ref{fig:planes}(a,b) the full flow field is well reconstructed at this distance from the
wall. The accuracy decays quickly with $y$, especially for $w$. At $y^+ = 40$ (corresponding
to the second row), the $R_{ww} = 0.5$ isoline only encloses wavelengths two to three times
longer and wider than those at $y^+ = 20$. Figure \ref{fig:spectra} contains data at two
Reynolds numbers. At the wall distances in the first two rows, which are in the buffer
layer, the error isolines collapse well in wall units. In this range, the widest
reconstructible spectral range is that of $v$, but, because the large scales of $v$ contain
much less energy than those of $u$, the latter is the best reproduced variable in terms of
energy. These relations persist up to $y^+ \approx 60$.

The trend to poorer reconstruction for larger $y$ persists as we move into the logarithmic
layer, as shown by the last two rows in figure \ref{fig:spectra}. They are normalised in
outer units and also collapse reasonably well for the two Reynolds numbers in the figure.
The spectral region that can be well reconstructed is different for the four variables. In
general, $w^\dagger$ has the narrowest spectral range, and $v^\dagger$ and $u^\dagger
v^\dagger$ the widest one, although it is interesting that the Reynolds product, $u^\dagger
v^\dagger$, can be reconstructed over a wider range that either $u^\dagger$ or $v^\dagger$.
Note that the spectral error only makes sense within the non-zero part of the
spectra, since otherwise \r{eq:R} is essentially undetermined.

The wavelengths of the error isolines above the buffer layer are approximately proportional
to their distance from the wall, confirming that the eddies that are `attached' in this
sense are self-similar \cite{tow:61}. This is illustrated in figure \ref{fig:spectra_sca},
where the reconstruction boundaries are shown at different heights, scaled both with $h$ and
with $y$. The good scaling with $y$ is evident in the lower row of figures. The spanwise
dimensions of the reconstructed $u^\dagger$, $v^\dagger$, and $u^\dagger v^\dagger$ are
$\lambda_z \gtrsim 2y$, which is in fair agreement with the dimensions of the attached
`quadrant' (Q-)eddies isolated in the logarithmic region by thresholding $uv$
\cite{lozano-Q}. The streamwise scales that can be reconstructed differ among variables. The
longest one is $u^\dagger$, probably because \textcolor{refcolor1}{the shorter features of $u$ are less energetic},
but $v^\dagger$ and $u^\dagger v^\dagger$ can be reconstructed down to $\lambda_x\approx
5y$, which is again in reasonable agreement with the dimension of Q-structures in
\cite{lozano-Q}. In any case, while the relation between different variables appears robust,
the absolute limits should not be taken too seriously. The 25\% error isolines in the lower
row of figure \ref{fig:spectra_sca} show that the limits depend on the chosen error
threshold.

As mentioned above, the spanwise velocity component is the worst reproduced at all heights.
As seen in figures \ref{fig:spectra} and \ref{fig:spectra_sca}, its \textcolor{refcolor1}{wavelength} range is approximately
40\% smaller than for $u$ or $v$, although it is interesting to note that the instantaneous
reconstructed fields satisfy continuity everywhere \citep{adr:moi:88}. \textcolor{refcolor1}{Even though the observable wavelengths of $w$ are limited to longer (and wider) scales than those of $v$, the reconstructions of both velocity components contain similar kinetic energy, as these scales of $w$ are more energetic than the large scales of $v$.} We will see below
that $w$ is mostly associated with the shear at the wall, suggesting streamwise rollers or
vortices. If this were instead the case, each roller would have a lower and an upper layer of $w$ and,
while the lower layer would leave a clear wall footprint, the upper one would be harder to
reconstruct.

Figure \ref{fig:reduced} compares the overall performance of the reconstructions by means of the ratio
\begin{equation}
  \beta_{ab}(y) = \frac{\sum_{\forall k_x, k_z} a^\dagger (b^\dagger)^*(k_x,y,k_z)}
  {\sum_{\forall k_x, k_z} ab^*(k_x,y,k_z)},
\end{equation}
between the energy of the reconstructed and original fields. The solid lines in the figure
show that reconstruction recovers more than 50\% of $u^2$ and $uv$ below $y/h\approx 0.2$,
but that the cross-flow velocities are only reconstructed to that accuracy within the buffer
layer, $y^+\lesssim 100$. All the reconstructions degrade above $y/h\approx $0.2-0.3.

\subsection{Incomplete wall data}\la{sec:onevar}

Not all observables contribute equally to the reconstruction. Their relative contribution is
different for each velocity component, and changes between the buffer and
the logarithmic layer. In this section, we explore these differences by repeating the
previous study using only one observable at a time\textcolor{refcolor1}{, as detailed in the discussion of \eqref{eq:lse0}}. The restricted reconstructed velocity
fields are denoted as ${a}^\dagger_p$, ${a}^\dagger_{u_y}$ and ${a}^\dagger_{w_y}$, where
$a$ stands for the variable being estimated, and the subindex is the observable. 

\begin{figure}
\centering
\includegraphics[width=\textwidth,clip]{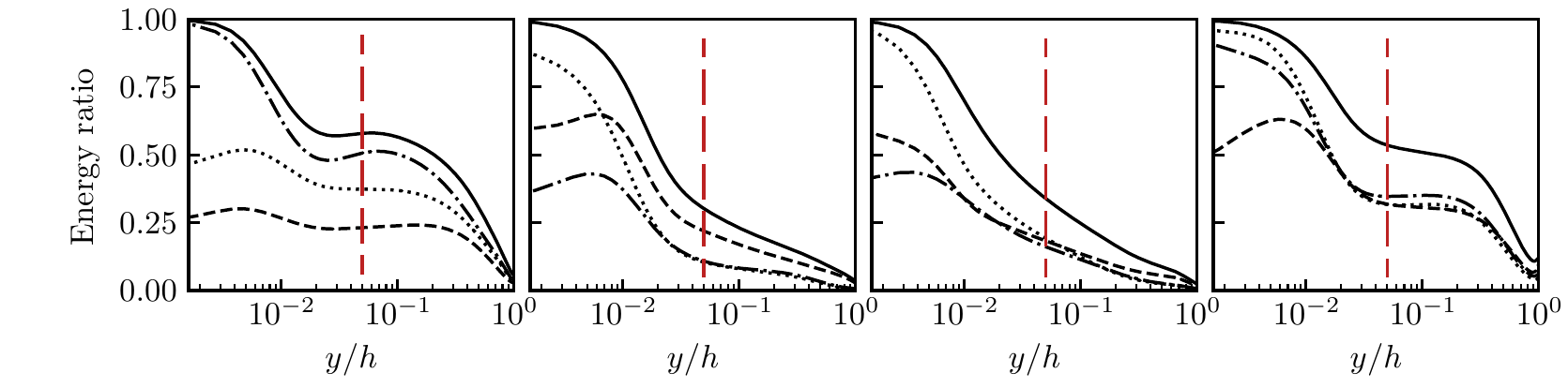}%
\mylab{-11.9cm}{3.3cm}{(a)}%
\mylab{-8.4cm}{3.3cm}{(b)}%
\mylab{-4.8cm}{3.3cm}{(c)}%
\mylab{-1.0cm}{3.3cm}{(d)}%
\\%
\includegraphics[width=\textwidth,clip]{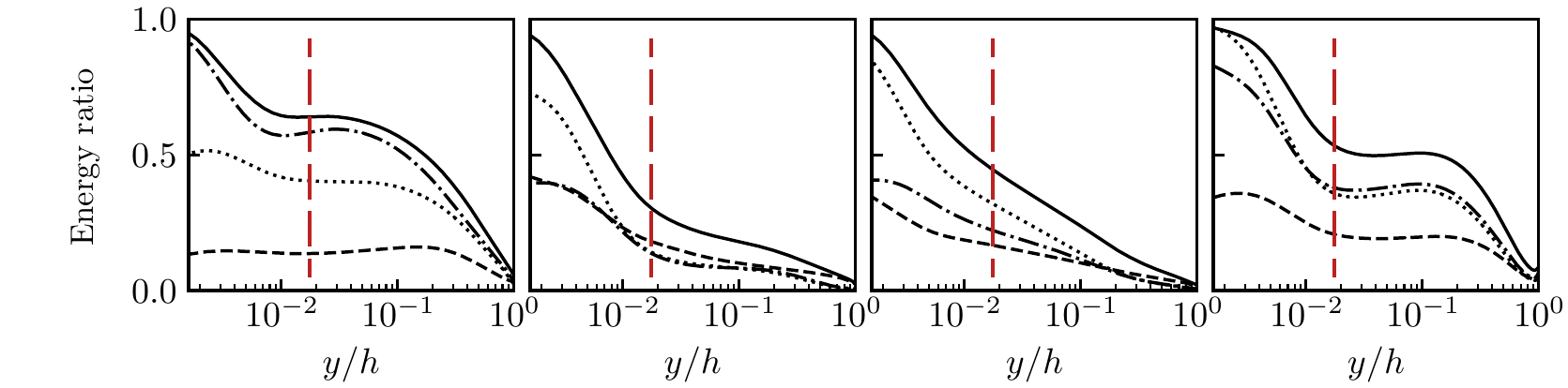}%
\mylab{-11.9cm}{3.3cm}{(e)}%
\mylab{-8.4cm}{3.3cm}{(f)}%
\mylab{-4.8cm}{3.3cm}{(g)}%
\mylab{-1.0cm}{3.3cm}{(h)}%
\caption{Reconstructed energy ratio with single-observable models for F2000 (top) and F5300
(bottom). The lines are: \solid, ${a}^\dagger$; \chndot, ${a}^\dagger_{u_y}$; \dotted
${a}^\dagger_{w_y}$; \dashed, ${a}^\dagger_p$. From left to right, $uu$, $vv$, $ww$, $uv$.
The dashed vertical lines are $y^+ \approx 100$, approximately corresponding to the end of
the buffer layer.}
\label{fig:reduced}
\end{figure}

Figure \ref{fig:reduced} compares the performance of the different restricted
reconstructions with that of the full model. The reconstruction with the highest energy
fraction is $u^\dagger$. Not surprisingly, the most important observable is $\p_y u$ in this
case, and $u^\dagger_{u_y}$ closely approximates the full model. The spanwise shear, and
especially the pressure, have much less influence on this component. On the other hand, $v$
is, on average, mostly reconstructed from the pressure, which seems reasonable, except very
near the wall, where the spanwise shear dominates. This is probably associated with the
presence of the near-wall streamwise vortices. The correlation between the wall pressure and
$v$ was studied in some detail by \cite{VilaFlores:2018}. They find that the correlation is
almost always maximum at zero spanwise offset, but that it has weak side-lobes at $\Delta
z^+\approx \pm 50$ when $v$ is in the buffer region, consistent with the known structure of
the streamwise vorticity in the sublayer \cite{kim:moi:mos:87}. The spanwise velocity is
mostly controlled by the spanwise shear near the wall, but all observables have a similar
influence  on it farther away. As we have already discussed, the reconstruction of this variable is
generally poor.

On the other hand, the tangential Reynolds stress, $u^\dagger v ^\dagger$, is fairly well
reconstructed up to $y/h \approx 0.3$. It is intriguing that $u^\dagger v ^\dagger$, behaves
better than $v^\dagger$, which is one of its factors, but we already mentioned when
discussing figure \ref{fig:streaks} that the reason is that the large $v$-structures that
can be reconstructed from the wall are precisely those associated with the large streaks of
the streamwise velocity. The relevant observables for  $u^\dagger v ^\dagger$ are
intermediate between those for $u^\dagger$ and $v^\dagger$. In the buffer layer, the two
shears are dominant, while the pressure is not. Farther away from the wall, the three
observables are comparable. Note that the good behaviour of this variable is encouraging from the
point of view of observing dynamics from the wall.

\begin{figure}
  \vspace*{3ex}%
  \centering
\includegraphics[width=.99\textwidth,clip]{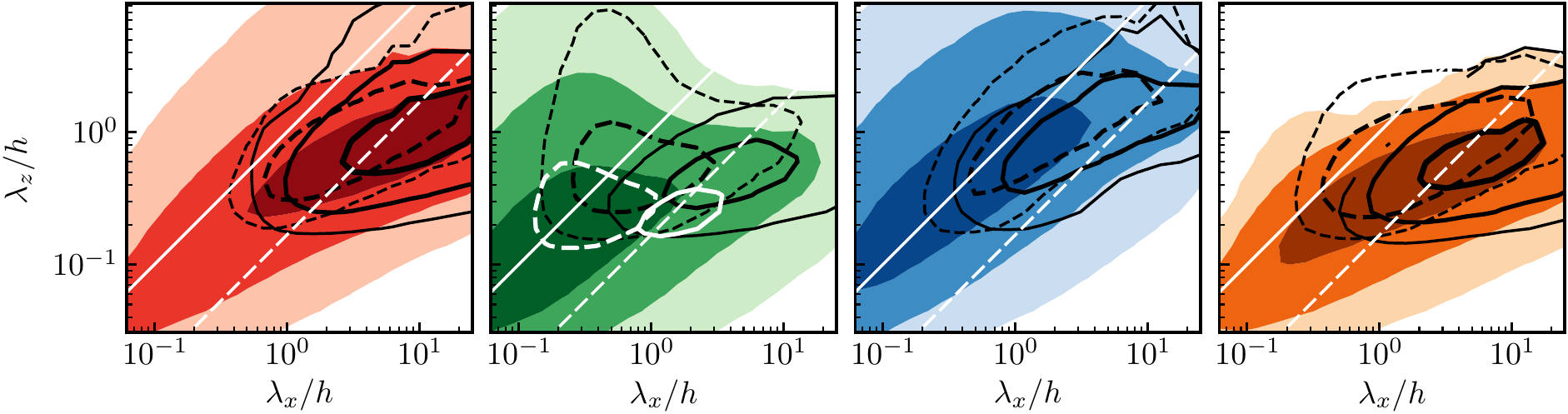}%
  \mylab{-13.5cm}{4.45cm}{(a)}%
  \mylab{-9.7cm}{4.45cm}{(b)}%
  \mylab{-5.9cm}{4.45cm}{(c)}%
  \mylab{-2.1cm}{4.45cm}{(d)}%
  \vspace{5ex}
  \includegraphics[width=\textwidth,clip]{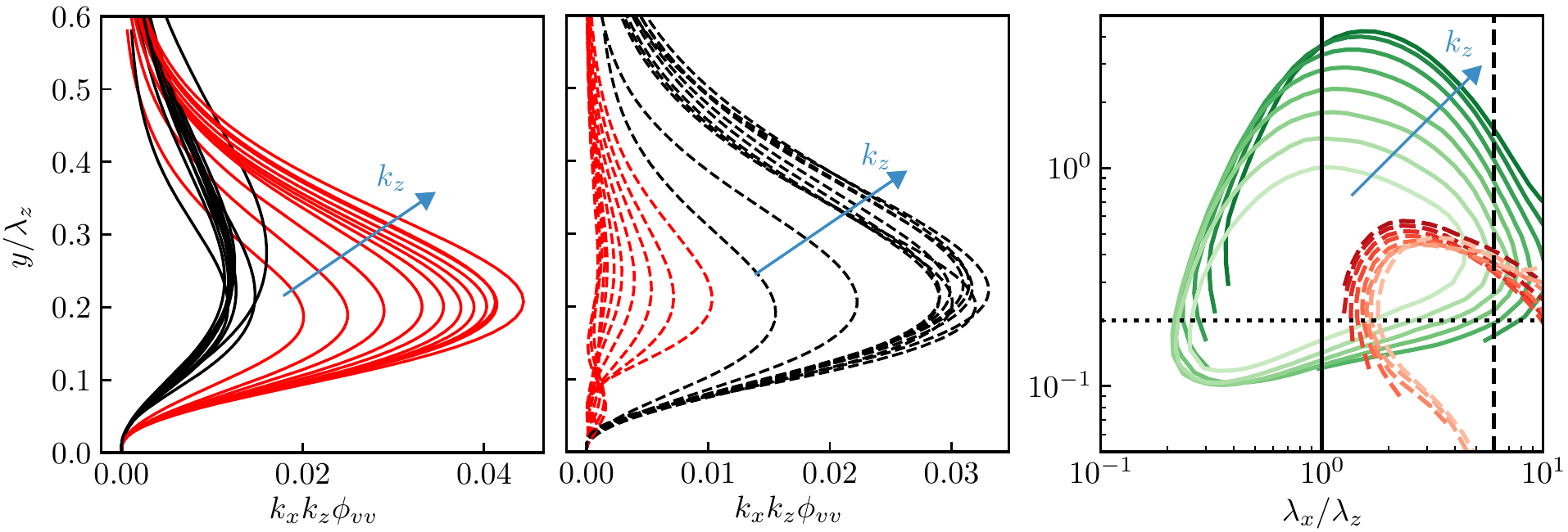}%
  \mylab{-13.3cm}{5.6cm}{(e)}%
  \mylab{-8.5cm}{5.6cm}{(f)}%
  \mylab{-3cm}{5.6cm}{(g)}%
\caption{(a-d) Premultiplied spectra and cospectrum at $y/h = 0.1$ for F5300. Shaded, spectrum of the
original fields; \solid, spectrum of the reconstruction using only the streamwise shear;
\dashed, using only the pressure. The contour levels for all the spectra are 1.5\%, 10\% and
50\% of the maximum of the corresponding true spectrum. The white contours in (b) are
reconstructions at $y/h = 0.05$.
The diagonal lines are: \solid, $\lambda_x=\lambda_z$; \dashed, $\lambda_x=6\lambda_z$. 
(a) $k_xk_z\phi_{uu}$. (b) $k_xk_z\phi_{vv}$. (c) $k_xk_z\phi_{ww}$. (d) $-k_xk_z\phi_{uv}$. 
(e-f) Premultiplied spectra of $v^\dagger_p$ (red) and $v^\dagger_{u_y}$ (black). 
(e) Modes along $\lambda_x = \lambda_z$. (f) Along $\lambda_x = 6\lambda_z$.
(g) Streamwise sections at different $\lambda_z\in (0.1-0.84)$, from dark to light, of: 
\solid, $k_xk_z\phi_{vv}^+ = 0.05$;  \dashed,  $k_xk_z\phi_{uu}^+ = 0.3$. 
The straight lines are, \solid, $\lambda_x=\lambda_z$; \dashed, $\lambda_x=6\lambda_z$; 
\dotted, $y = 0.2\lambda_z$.
}
  \label{fig:reduced_spe}
  \end{figure}

The reconstructions for the two Reynolds numbers agree reasonably well in inner units close
to the wall $(y^+<100)$, and in outer units far from it. The obvious exception are the
reconstructions based only on the pressure in figure \ref{fig:reduced}, which contain
considerably less energy at the higher Reynolds number. This trend is most pronounced for
$u^\dagger_{p}$ and, consequently, for $u^\dagger_{p}v^\dagger_{p}$, for which the reconstructed
energy is 50\% smaller for F5300 than for F2000. The degradation is restricted to the reconstruction from the
pressure, and does not affect either the full model or the reconstructions in terms of the shears. Inspection
of the spectra of $u^\dagger_{p}$ and $v^\dagger_{p}$ at the two Reynolds numbers reveals that
most of the energy missing from F5300 resides in relatively 
elongated scales, which we will see below to be dominated in the full model by the shear (see
figure \ref{fig:reduced_spe}). A possible explanation for this behaviour follows from the
decomposition of the pressure at the wall into an `inertial' component,
\begin{equation}
  \nabla^2p_I = -\nabla\cdot(u\cdot\nabla u),\qquad \partial_y p_I\big|_{y=0} = 0,
\end{equation}
and a `Stokes' one \cite{kimpres89},
\begin{equation}
  \nabla^2p_s = 0,\qquad \partial_yp_s\big|_{y=0} = \nu\partial_{yy}v,
\end{equation}
\textcite{encinar:ctr18} showed that the latter can be computed from
the wall shears using continuity,
\begin{equation}
\hp^+_s(0) = -\ii \left [ k_x^+ \hu_y(0)^+ + k_z^+ \hw_y(0)^+\right ]/|k|^+.
\la{eq:pstokes} 
\end{equation}
It can be shown that the ratio of between $\hp_s$ and $\hp_I$ is approximately independent
of the Reynolds number when the wavenumbers are scaled in wall units, but that it decays as
$\lambda^+_x$ increases along the diagonal $\lambda_z=\lambda_x$, which contains the cores
of the velocity spectra (see figure \ref{fig:reduced_spe}), essentially because
$k_xk_z\phi_{pp}(0)$ is located along this diagonal, but the two shears are not (figure
\ref{fig:disip}b). The result is that, as $Re_\tau$ increases and we use our estimation to
reconstruct $u^\dagger_p$ at a fixed $y/h$, and therefore at increasing $\lambda^+_x\sim
y^+=Re_\tau y/h$, the importance of the Stokes component decreases. It follows from
\r{eq:pstokes}, that $p_s$ is actually part of the shears, but there is no reason why the
particular combinations of shears in \r{eq:pstokes} is in any way optimal to reconstruct the
velocities. The pressure-only reconstructions are therefore contaminated by $p_s$, and
should asymptote to a `pure' reconstruction in terms of $p_I$ as $Re_\tau$ increases. In
this view, the `purest' $u^\dagger_p$ is the one for F5300, and the apparently higher energy
fraction in F2000 is accidental. The full model does not have this problem, because it uses
information from the two shears to compensate the spurious effect of $p_s$. As a
consequence, its performance in figure \ref{fig:reduced} does not change with $Re_\tau$.

Figure \ref{fig:reduced} is contaminated below $y^+\approx 100$ (indicated by
a vertical red line) by the relatively low resolution of the data stored for F2000 and
F5300, and we used the higher-resolution smaller box S1000 to estimate the effect of the
contamination. The only difference found was a better performance of $v^\dagger_{p}$ and $w^\dagger_{p}$ (but not of $u^\dagger_{p}$), most probably because of a
better estimation of the streamwise vortices of the buffer layer. This idea is supported by the high value of the cross-correlation between $w_y$ and $p$ at the wall, which is close to unity for the small scales, and the fact that $w_y$ is the best predictor of the cross velocities in the buffer region.

Figure \ref{fig:reduced_spe}(a-d) displays the spectral distribution of the energy
reconstructed by the different single observables. In most cases, the wall pressure and the
two shears predict similar spectral regions, comparable to those of the the full
reconstructions in figure \ref{fig:spectra_sca}, but the behaviour of $v^\dagger$ in the
logarithmic layer is interesting, because different observables dominate in different
spectral ranges.

The spectrum of $v$ has a wide and a narrow component in the logarithmic layer
\citep{jimhoy08}, which appear in figure \ref{fig:reduced_spe}(b) as the two `horns' on the
upper right-hand corner of the spectrum. The spanwise wall shear contributes most to the estimation of
the longer component, and the pressure dominates the shorter one, suggesting, respectively,
streamwise rollers and compact sweeps and ejections. The reconstruction from the pressure is
centred on equilateral wavenumbers, whereas the reconstruction from the streamwise shear
dominates at wavelengths elongated in the streamwise direction. Only $v^\dagger_{u_y}$ is
shown in figure \ref{fig:reduced_spe}, but the effect on $v^\dagger$ of the two shears  is
similar above $y^+\approx 100$.

That this separation among $v$-structures is not restricted to a particular distance from the wall is shown in
figure \ref{fig:reduced_spe}(e,f), which displays profiles of the energy spectrum of
$v^\dagger_p$, in red, and of $v^\dagger_{u_y}$, in black, at various points along the two
diagonals in figure \ref{fig:reduced_spe}(b). Along the equilateral solid diagonal for which
the pressure reconstruction is dominant, $\lambda_x=\lambda_z$, the spectrum reconstructed
from the pressure is approximately three times more intense than that of $v^\dagger_{u_y}$
(figure \ref{fig:reduced_spe}e), and the opposite is true for the longer structures in
figure \ref{fig:reduced_spe}(f), $\lambda_x=6\lambda_z$. It is interesting that the peak of
the reconstructed spectra is at $y/\lambda_z\approx 0.2$ for all the wavelengths for which
this peak is in the logarithmic layer. When a similar plot is drawn for the true $v$, the peak is at
$y/\lambda_z\approx 0.3-0.5$. Reconstruction from either observable only recovers structures
closer to the wall than the average for the real flow field, presumably corresponding to the
`roots' of taller attached eddies.

\begin{figure}
  \centering
  \includegraphics[width=0.95\textwidth]{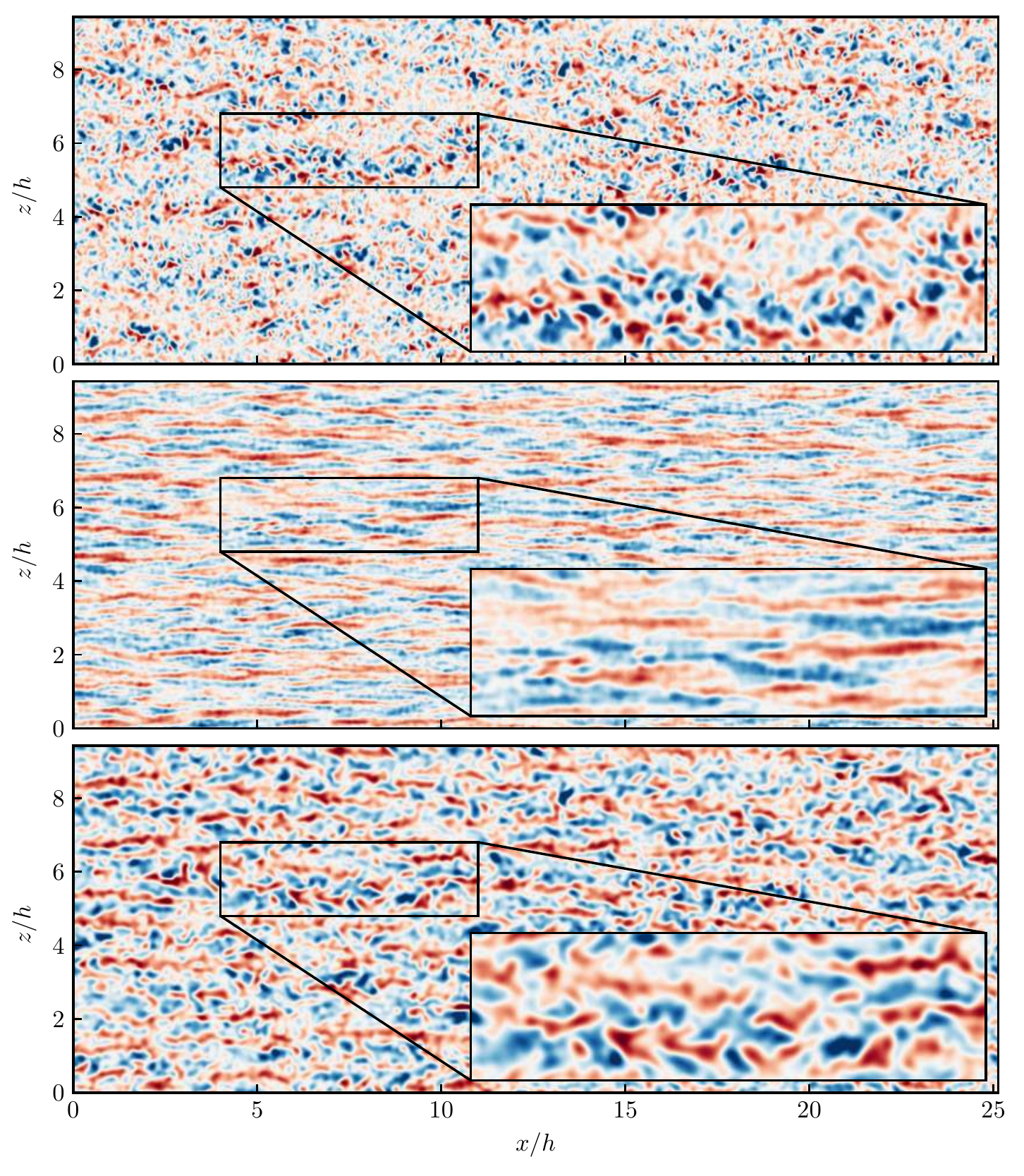}
\caption{Reconstruction of the wall-normal velocity at $y/h=0.1$ in F5300. Top, using only
the pressure, $v^\dagger_p$. Middle, using only the wall shear, $v^\dagger_{u_y}$. \textcolor{refcolor2}{Bottom, true velocity field, filtered with \eqref{eq:G} and $\Delta_x \times \Delta_z (2\times 2)y$. The shading goes from $-u_\tau$ to $u_\tau$.}
  }
  \label{fig:mystery_planes}
\end{figure}
 
Further insight into the difference among the two reconstructions is provided by figure
\ref{fig:reduced_spe}(g), which contains streamwise sections of the premultiplied spectrum
of $u$, in red, and of $v$, in green\textcolor{refcolor1}{; represented against the wavenumber aspect ratio}. The reconstruction peak at $y/\lambda_z=0.2$ is marked by
the horizontal dotted line, and intersects the lower edge of the $v$-spectrum. The two
diagonals in figure \ref{fig:reduced_spe}(b) are represented by vertical lines in this plot\textcolor{refcolor1}{, as they contain wave numbers with constant aspect ratio}.
The solid vertical line is the equilateral diagonal, and crosses the centre of the
$v$-spectrum. This is the spectral location at which $v^\dagger$ is best reconstructed by
the wall pressure, and figure \ref{fig:reduced_spe}(g) suggests that it
contains $v$-structures whose lower edge interacts with the impermeability of the wall to
create a pressure peak\textcolor{refcolor1}{, as larger modes are damped faster when they approach the wall}. In contrast, the elongated diagonal $\lambda_x=6\lambda_z$,
represented in the figure by the dashed vertical line, crosses the peak reconstruction wall distance
\textcolor{refcolor1}{at the edge of} the $v$-spectrum, but near the centre of the spectrum of $u$. This
$v$-structures are too far from the wall to create an overpressure, but they can be detected from
the wall shear because they are associated with the streaks of $u$\textcolor{refcolor1}{, whose spectrum reaches the wall in this region}. Finally, figure
\ref{fig:mystery_planes} displays two $v^\dagger$ snapshots reconstructed from the pressure
(top), and from the wall shear (bottom). The difference in geometry is obvious, and the
lateral spacing between streaky $v$-structures of similar sign in the bottom snapshot
$(\Delta z/h\approx 0.6, \Delta z^+ \approx 3200)$ approximately agrees with the separation between the streamwise
velocity streaks at the distance from the wall of the figure (see the spectrum in figure
\ref{fig:reduced_spe}a). It is interesting that, although it follows from figure
\ref{fig:reduced}(h) that the total $uv$-energy reconstructed from the streamwise shear is approximately
twice larger than the one reconstructed from the pressure, the tangential Reynolds stress contained in intense structures \cite{lozano-Q} of
the two snapshots in figure \ref{fig:mystery_planes} are approximately equal (not shown), with similar
intensity and area fraction.

\subsection{Reconstruction in physical space}\la{sec:physical} 

\begin{figure}
  \centering
\includegraphics[width=.99\textwidth]{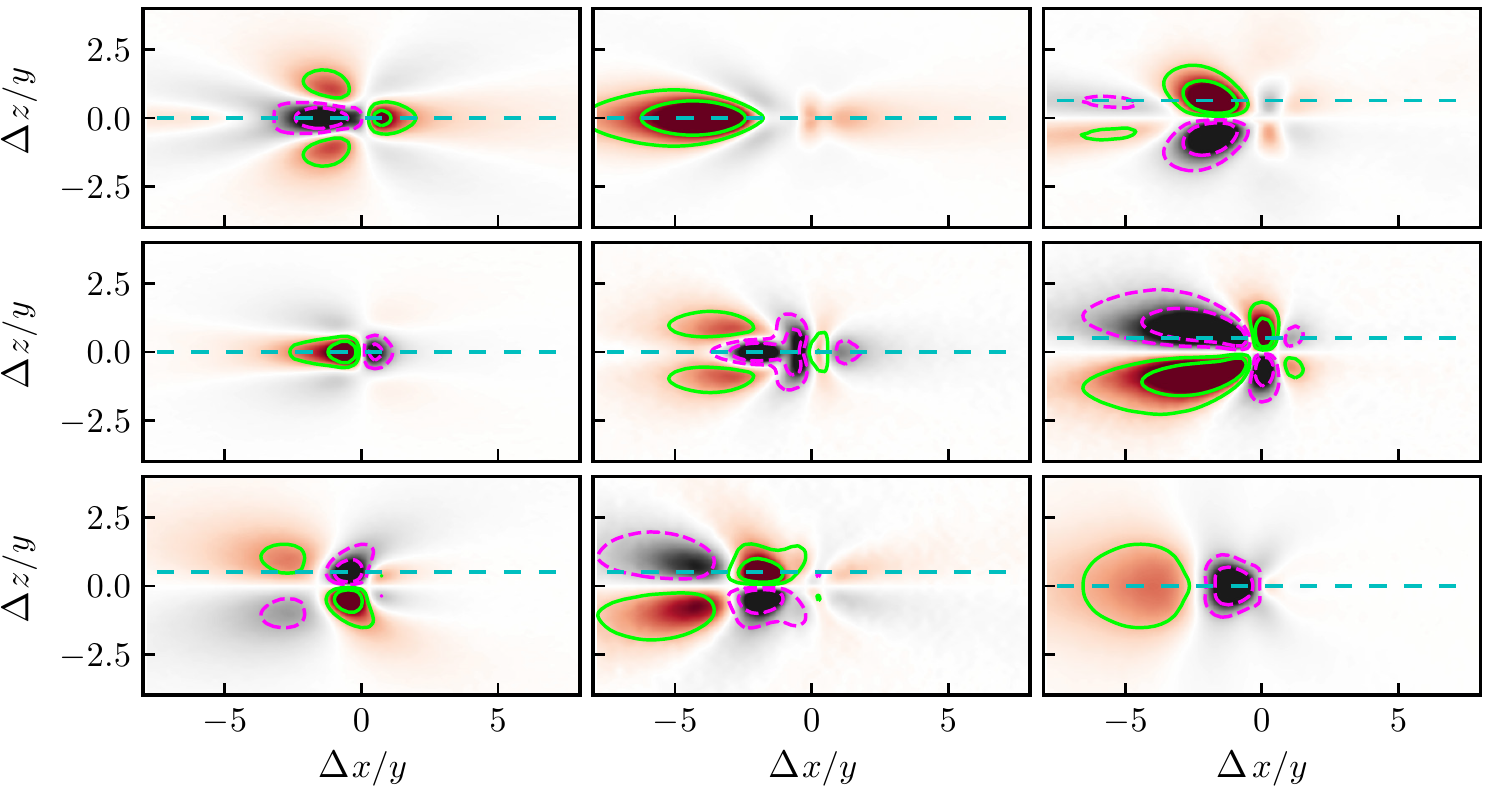}%
\mylab{-0.66\textwidth}{0.48\textwidth}{(a)}%
\mylab{-0.36\textwidth}{0.48\textwidth}{(b)}%
\mylab{-0.06\textwidth}{0.48\textwidth}{(c)}%
\mylab{-0.66\textwidth}{0.325\textwidth}{(d)}%
\mylab{-0.36\textwidth}{0.325\textwidth}{(e)}%
\mylab{-0.06\textwidth}{0.325\textwidth}{(f)}%
\mylab{-0.66\textwidth}{0.17\textwidth}{(g)}%
\mylab{-0.36\textwidth}{0.17\textwidth}{(h)}%
\mylab{-0.06\textwidth}{0.17\textwidth}{(i)}%
\caption{Components of the operator $\tL_s$ for F5300. The rows are the operator for $u$,
$v$ and $w$, from top to bottom. The columns are the components 
corresponding to $p$, $u_y$ and $w_y$, from left to right. Shaded contours are for
$y/h=0.1$, positive in red and negative in black. Lines are for
$y/h=0.05$, positive in green and negative in magenta. The contour levels are
$[-0.5, -0.2, 0.2, 0.5]$ of the maximum absolute value at each height. The dashed
lines indicate the vertical sections in figure \ref{fig:LSE_xy}.
}
  \label{fig:LSE_xz}
\vspace*{2mm}%
\includegraphics[width=.99\textwidth]{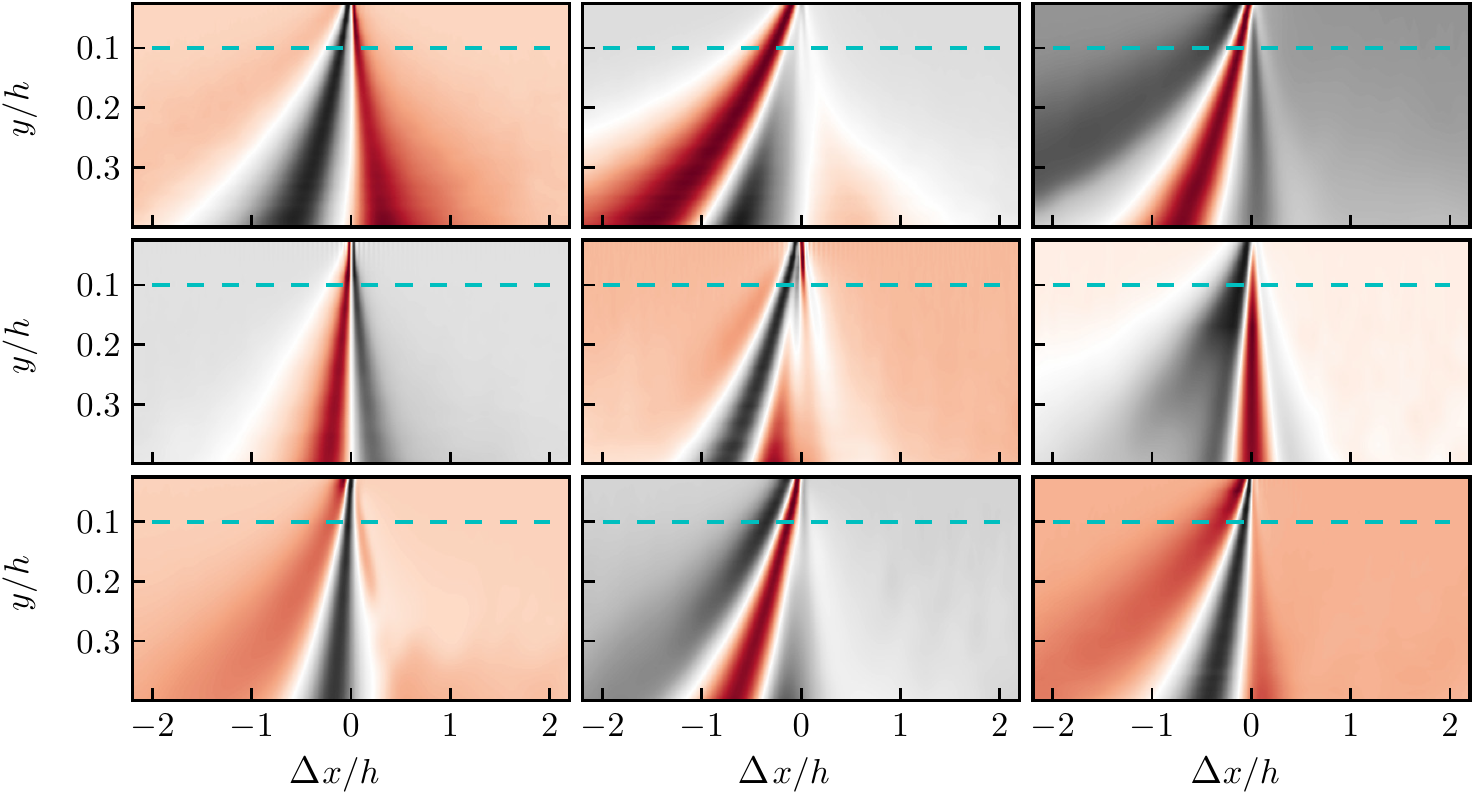}%
\mylab{-0.66\textwidth}{0.51\textwidth}{\colorbox{white}{(a)}}%
\mylab{-0.36\textwidth}{0.51\textwidth}{\colorbox{white}{(b)}}%
\mylab{-0.06\textwidth}{0.51\textwidth}{\colorbox{white}{(c)}}%
\mylab{-0.66\textwidth}{0.35\textwidth}{\colorbox{white}{(d)}}%
\mylab{-0.36\textwidth}{0.35\textwidth}{\colorbox{white}{(e)}}%
\mylab{-0.06\textwidth}{0.35\textwidth}{\colorbox{white}{(f)}}%
\mylab{-0.66\textwidth}{0.19\textwidth}{\colorbox{white}{(g)}}%
\mylab{-0.36\textwidth}{0.19\textwidth}{\colorbox{white}{(h)}}%
\mylab{-0.06\textwidth}{0.19\textwidth}{\colorbox{white}{(i)}}%
\caption{As in figure \ref{fig:LSE_xz}. Streamwise sections: $z = 0$ for symmetric
components, and at the location of the maximum absolute value at each height for
antisymmetric ones. The dashed line is $y/h = 0.1$. The shading is scaled with the maximum absolute value at each height.}
  \label{fig:LSE_xy}
\end{figure}

Although the spectral expression of the reconstruction operator discussed in the previous
section is very useful in understanding the behaviour of the different flow scales, its
physical-space counterpart defined in \r{eq:last_lse} and \r{eq:cont_lse} gives a more
intuitive representation of the operation that is actually being performed. It is probably
also easier to generalise to wall-bounded flows besides the channel. We saw in
\S\ref{sec:methods} that the LSE operator is essentially a two-point correlation scaled with
the cross-correlation tensor of the observables at the wall, and both correlations are
likely to be reasonably independent of the type of flow in the near-wall region
where the reconstruction is effective. For example, this is the case for boundary layers and
channels below $y/h\approx 0.1$, although they differ farther from the wall
\cite{sil:jim:14}.

Figures \ref{fig:LSE_xz} and \ref{fig:LSE_xy} show sections of the physical-space
reconstruction operator $\tL_s^{(a)}$ for the three velocity components, $a=\{u, v, w\}$,
and the three observables, $s=\{p,u_y, w_y\}$. Note that, although $\tL$ is formally a
function of the two wall-parallel coordinate increments, $\Delta x = x-x'$ and $\Delta z =
z-z'$, and of the reconstruction height $y$, their role is different. The operator is always
two-dimensional in $(\Delta x,\Delta z)$, and acts as a convolution kernel on $E(x',z')$ at
the wall. The wall-normal coordinate $y$ is a parameter, and the operators for the
reconstruction at different wall distances are, in principle, unrelated to each other. To
emphasise this, the coordinates in these two figures have been inverted. The operator for
one wall distance is given by the section of the three-dimensional function $\tL(\Delta x,
y, \Delta z)$ at one $y$, as in figure \ref{fig:LSE_xz}, and $\Delta x<0$ implies that the
convolution kernel is acting on the wall upstream of the reconstruction point.

The wall-parallel reconstruction kernels in figure \ref{fig:LSE_xz} show that the main
reason for the lack of small scales in the reconstructions is that operators act as
smoothing filters. All kernels have positive and negative parts, and their relative
arrangement can partly be predicted from symmetry considerations. For example,
$\tL^{(u)}_{u_y}$ in figure \ref{fig:LSE_xz}(b) is symmetric with respect to $\Delta z=0$
because both $u$ and $u_y$ are symmetric with respect to $\Delta z\to -\Delta z$, but
$\tL^{(u)}_{w_y}$ in figure \ref{fig:LSE_xz}(c) is antisymmetric because $u$ is symmetric
and $w_y$ is antisymmetric. This has to be taken into account in interpreting figure
\ref{fig:LSE_xz}, because the shape of some of the kernels suggests that the relevant
variable is not the observable but some of its derivatives. For example, the
positive-negative alternation along the $\Delta x$ axis\textcolor{refcolor1}{, centred at
$\Delta x = 0$,} of $\tL^{(u)}_{p}$ in figure \ref{fig:LSE_xz}(a) probably implies that the
relevant observable is not $p$ but $\p_x p$, because there is no statistical symmetry to
$\Delta x\to -\Delta x$\textcolor{refcolor1}{, and thus $\tL^{(u)}_{\p_x p}$ should be
maximum at $\Delta x=0$. A similar argument can be made from the Fourier counterpart of
\eqref{eq:lse0}, which, for a single Fourier mode is
\begin{equation}
  \langle \hat p^* \hat p \rangle \hat L^{(u)}_{p} = \langle \hat p^* \hat u \rangle.\label{eq:pfou}
\end{equation}
This equation is related to the one for the estimation of $u$ from
$\widehat{\partial_x  p} = ik_x \hat p$,
\begin{equation}
  -k_x^2\langle \hat p^* \hat p \rangle \widehat{L}^{(u)}_{\p_x p} = ik_x\langle \hat p^* \hat u \rangle,
\end{equation}
which is identical to \eqref{eq:pfou} except for a factor multiplying each side of the
equation. The left-hand side its multiplied by $-k_x^2$, which is a even factor. This is the
only possibility, because the first factor represents the autocorrelation of $\partial_x p$,
which by definition is a symmetric function. On the other hand, the right-hand side gets
multiplied by the odd factor $ik_x$, which implies that, for a dominantly skew-symmetric
$\tL^{(u)}_{p}$, such as the one in
figure \ref{fig:LSE_xz}(a), the operator  $\tL^{(u)}_{\p_x p}$ should be approximately symmetric along the streamwise direction, peaking at $\Delta x\approx 0$}.
The similar alternation in $\Delta z$, which would suggest $\p_{zz} p$, may be due to the
spurious symmetry of the correlation along the $\Delta z$ axis, and could equally be
interpreted as a statistical distortion of $\p_z p$ in which 50\% of the events fall to
either side of the reconstruction point. For a discussion of symmetry artifacts of LSE in
the context of the local reconstruction of the buffer layer, see \cite{stretch90}.

The evolution of the reconstruction kernels with the distance from the wall of the target
point is displayed in figure \ref{fig:LSE_xy}. Symmetric kernels are plotted at the symmetry
plane, $\Delta z=0$, but antisymmetric ones, for which this plane is identically zero, are
plotted at the $\Delta z$ plane for which the kernel is most intense. A general property of all
the kernels is that they become wider as $y$ increases, and that the growth is almost
linear, suggesting self-similarity with respect to $y$. In fact, the wall-parallel plot in
figure \ref{fig:LSE_xz} is scaled with $y$, and includes two wall distances, whose kernels
collapse well in this representation. Another property of most kernels in figure
\ref{fig:LSE_xy} is that they tilt forward, implying that the flow is reconstructed from
upstream wall variables. Reference \cite{VilaFlores:2018} found that the correlation of $v$
with the pressure at the wall is antisymmetric in $\Delta x$, with both upstream and
downstream lobes, but figure \ref{fig:LSE_xy}(d) shows that the corresponding kernel,
$L^{(v)}_p$, is one of the few that is not tilted in $\Delta x$. This approximately applies
to all the kernels based on the wall pressure, but those based on the shears are tilted by
an angle that depends on the particular case, but which is of the order of
10\degree--20\degree\ to the wall, similar to other structural angles measured for $u$ and
$w$ in the logarithmic layer of shear flows \cite{sil:jim:14}. \textcolor{refcolor1}{The disparity between the tilting of the shear and pressure kernels can be explained if we consider the processes generating the signals at the wall. If an attached eddy of size $O(l)$ is centred at a distance to the wall of $y_l \sim l$, the time it takes for a change of its dynamics to affect the shear at the wall should be of the order of $t_l \sim y_l/u_\tau \sim l/u_\tau$, equivalent to his local turnover time. Coincidentally, this time is related to the shear time by the Corrsin shear parameter, which is of the order of 10 for the logarithmic region of channel flows \cite{jim13_lin}. This implies that the eddy is advected downstream during the time it takes for the information to reach the wall. In contrast to that, the pressure of incompressible fluids instantly transmits  sudden changes globally, generating a signal at the wall with no offset.}

\subsection{Off-design behaviour}\la{sec:off} 
  
\begin{figure}
\vspace*{3ex}%
\centering
\includegraphics[width=0.99\textwidth,clip]{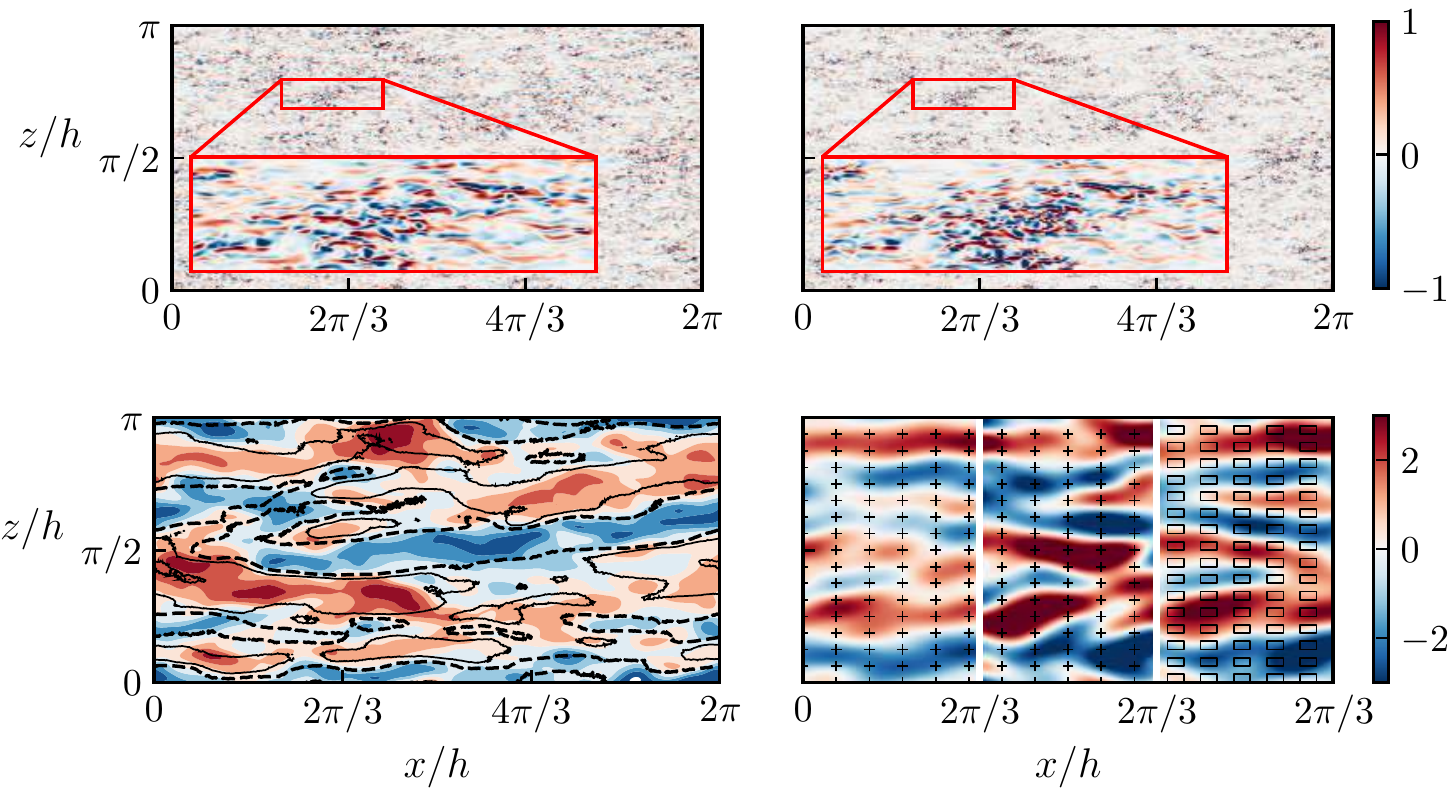}
\mylab{-0.715\textwidth}{0.53\textwidth}{(a)}%
\mylab{-0.283\textwidth}{0.53\textwidth}{(b)}%
\mylab{-0.715\textwidth}{0.265\textwidth}{(c)}%
\mylab{-0.41\textwidth}{0.265\textwidth}{(d)}%
\mylab{-0.283\textwidth}{0.265\textwidth}{(e)}%
\mylab{-0.17\textwidth}{0.265\textwidth}{(f)}%
\caption{Snapshots of different reconstructions. 
(a) S2000, $v^+$ at $y^+\approx 10$. 
(b) S2000, $(v^\dagger_{\text{S1000}})^+$ at $y^+\approx 10$. 
(c) S2000 at $y/h \approx 0.1$. Shaded, $u^+$; \full, $(u^\dagger_{\text{S1000}})^+=1$; \dashed, $(u^\dagger_{\text{S1000}})^+=-1$. S2000 is filtered with $G$ and $\Delta_x \times\Delta_z = (0.4\times 0.2)$. 
(d-f) S1000, $(u^\dagger)^+$ at $y/h \approx 0.1$,
with information limited to discrete sensors (black crosses).
(d) Fully dealiased. (e) No dealiasing. (f) Sensors averaged over 25\% of the wall surface.
}
\label{fig:scaling}
\end{figure}

Although the reconstruction operators are $\Rey$-dependant, they are obtained from 
two-point correlation tensors which are known to collapse in inner units close to the wall,
and in outer units farther away \citep{sil:jim:14}. This scaling extends to the linear
estimators, allowing operators developed from DNS to be used at a different Reynolds number
than the one at which they were obtained. Figure \ref{fig:scaling}(a,b) presents a snapshot 
of the true $v$ and of the reconstructed $v^\dagger$ in the buffer layer of S2000, and figure \ref{fig:scaling}(c)
presents a similar reconstruction of $u$ in the logarithmic layer. Both reconstructions are
performed using an operator computed from S1000. For the buffer-layer reconstruction $(y^+
\approx 10)$, the observables, velocities and operators have been scaled in inner units,
and, for the logarithmic region $(y/h \approx 0.1)$ they have been scaled in outer units.
Although there is some degradation of accuracy with respect to the reconstructions at the
nominal Reynolds number in figure \ref{fig:planes}, the result is still reasonable. The
worst errors are in scales that are not contained in the rescaled operators. For example,
for computational boxes with similar dimensions in outer units, operators obtained at lower
Reynolds numbers lack some of the longest scales when scaled in wall units. Similarly, some
of the smallest scales are missing from those operators when scaled in outer units. The
reconstructions in figure \ref{fig:scaling}(b,c) have been correspondingly mildly filtered to
avoid aliasing artefacts. The reconstruction operators are models of the structure of the
flow, and the success of the rescaling supports the intuition that the attached turbulent
structures near and far from the wall approximately scale in inner and outer units,
respectively.

\subsection{Practical application with limited input data}\la{sec:lim} 
Lastly, we explore some of the issues pertaining to the practical application of this
methodology. Figure \ref{fig:scaling}(d-f) shows $u^\dagger$ at $y/h = 0.1$ for S1000,
reconstructed from wall observations which are assumed to be only known at a coarse grid of
sensors spaced by $\Delta_x/h = 0.2$ and $\Delta_z/h = 0.1$, represented in the figure by
black crosses. Sampling has the effect of limiting the spatial resolution of the observables
to wavelengths longer than the Nyquist limit $(2 \Delta_x, 2\Delta_z)$, and of aliasing the
information of the shorter wavelengths into longer ones that fall within this limit
\citep{canuto88}. If the information can be properly dealiased by filtering the short
wavelengths before sampling, the only effect is the loss of resolution, as in figure
\ref{fig:scaling}(d). Because of the linearity of the reconstruction and the homogeneity in the
wall-parallel planes, this is equivalent to {\em a posteriori} filtering of a reconstruction
performed at full resolution. Unfortunately, this is usually impossible, because the full
measurements are unavailable. Figure \ref{fig:scaling}(e) shows the worst-case scenario in
which `point' sensors are used without dealiasing. The reconstruction is heavily distorted,
mostly in the form of higher intensities, and the effect is more severe than what could be
expected from simple aliasing of the wall observables. The reason is that the latter contain a lot of small
scales which should have no effect on reconstructions far from the wall, but which are
aliased into larger scales and amplified by the linear operator into large spurious features
of the reconstruction. A more realistic scenario is presented in \ref{fig:scaling}(f), in
which the sensors are modelled as finite rectangles with size $\Delta_x/h = 0.1$ and
$\Delta_z/h = 0.05$, and assumed to provide values averaged over their surface. This {\em a
priori} filtering does not fully eliminate the aliasing error, but reduces it significantly
by reducing the short-wavelength content of the wall observables. In a real experimental
application, where the temporal evolution of the sensor signal is available, time filtering
can also be used to limit the aliasing error. The lifetimes \citep{lozano-time} and
convection velocities \citep{jcadvel} of the turbulent structures in the logarithmic layer
are known, and are typically significantly longer and faster than those of the scales
responsible for the aliasing of the wall signal. Time filtering of the signals at those
scales mimics streamwise averaging through Taylor's approximation, and supplements the
benign effect of non-zero sensor size.

\section{Conclusions}\label{sec:conclusions}

We have shown that the three velocity components in turbulent channels can be reconstructed
reasonably well below $y/h\approx 0.2$ with linear stochastic estimation, using only the
observed pressure and the streamwise and spanwise shear at the wall. Only eddies attached in
the sense of having sizes comparable to $y$ \cite{tow:56}, are reconstructed, but these are
found to contain more than approximately 50\% of the kinetic energy of $u$, and of the
tangential Reynolds stress, $uv$, in the logarithmic layer. The optimum reconstruction uses
the three wall observables, but not all them are equally important, and the relative balance
among them depends on the variable being reconstructed and on the distance from the wall. We
have proposed that this can be used to sharpen the definition of which velocity components
are attached to the wall in which spectral ranges, defining as attached those eddies that can be reconstructed from
wall information, regardless of the particular reconstruction method. The results differ
from the generally accepted ones in some cases. For example, the wall-normal velocity, which is usually
considered a detached variable because it is inhibited near the wall by the impermeability condition, is
linked to the wall by the pressure, for equilateral scales, and by the wall shears, for longer
scales in which $v$ is associated with the $u$-streaks. Interestingly, although the two sets
of attached $v$-structures defined in this way, are very different, with a different balance
of $u$ and $v$ components, both are active in the sense of generating a comparable amount of
tangential Reynolds stress.

Although most of the paper deals with mode-by-mode reconstructions in the Fourier
representation of the flow, we have shown that the expressions of the operators in physical
space gives interesting information on the flow physics. Thus, while the wall pressure
predominately reconstructs velocities immediately above the the observation point, the
effect of the shear is transmitted along lines inclined by 10\degree--20\degree\ from the
wall, comparable to previous observations of the structures of $u$ and $w$ in wall-bounded
flows.
 
We have also shown that the reconstruction operators obtained at one Reynolds number can be
adapted to a different one by proper scaling, making them potentially useful for control
strategies in which only wall observations are available, or, in turbulence physics, for the
study of the flow dynamics from the wall. They can be adapted to experimental set-ups in
which the instantaneous measurements are only available on a coarser grid of sensors, but
only after careful dealiasing of the observations. Although not explicitly shown, it can be inferred from figure \ref{fig:reduced} that the information provided by the two shears is equivalent in the logarithmic region, and a practical grid of sensors may obviate the spanwise shear with almost no penalty in the accuracy of the reconstructions. Note that this does not apply to the flow in the buffer layer, where the spanwise shear is the most relevant observable for the cross flow.

The reconstruction operators for the four cases considered will be made available at
\url{https://torroja.dmt.upm.es/channels/data/}.

\acknowledgments{
This work was supported by the European Research Council under the Coturb grant
ERC-2014.AdG-669505, and performed in part during the CTR 2018 Summer Program at Stanford
University, whose hospitality is gratefully acknowledged.
}
  

\bibliography{ctr18}

\end{document}